 \documentclass[preprint, superscriptaddress, preprintnumbers, amsmath, amssymb]{revtex4}

\allowdisplaybreaks
\allowdisplaybreaks[4]
\usepackage{CJK}
\usepackage{graphicx}% Include figure files
\usepackage{dcolumn}% Align table columns on decimal point
\usepackage{bm}% bold math
\usepackage[dvipdfm,
            pdfstartview=FitH,
            CJKbookmarks=true,
            bookmarksnumbered=true,
            bookmarksopen=true,
            colorlinks=true,
            pdfborder=001,
            linkcolor=blue,
            anchorcolor=blue,
            citecolor=blue
            ]{hyperref}

\begin{document}

\begin{CJK}{GBK}{song}

\title{Possible chiral doublets in $^{60}$Ni}

\author{J. Peng}\email{jpeng@bnu.edu.cn}

\affiliation{Department of Physics, Beijing Normal University,
Beijing 100875, China}

\author{Q. B. Chen}\email{qbchen@pku.edu.cn}

\affiliation{Physik-Department, Technische Universit\"{a}t
M\"{u}nchen, D-85747 Garching, Germany}

\date{\today}

\begin{abstract}

The open problem on whether or not the chirality exists in doublet
bands M1 and M4 in light-mass even-even nucleus $^{60}$Ni is studied
by adopting the recently developed fully quantal four-$j$ shells
triaxial particle rotor model. The corresponding experimental energy
spectra, energy differences between doublet bands, and the available
$B(M1)/B(E2)$ values are successfully reproduced. The analyses on
the basis of the angular momentum components, the azimuthal plots,
and the $K$-plots suggest that the chiral modes exist at $I\geq
12\hbar$ in doublet bands M1 and M4.

\end{abstract}

\maketitle

%%%%%%%%%%%%%%%%%%%%%%%%%%%%%%%%%%%%%%%%%%%%%%%%%%%%%%%%%%
%                    begin  introduction
%%%%%%%%%%%%%%%%%%%%%%%%%%%%%%%%%%%%%%%%%%%%%%%%%%%%%%%%%%

%\section{Introduction}

Nuclear chiral rotation is an exotic form of spontaneous symmetry breaking,
which exists only in nucleus with triaxial ellipsoidal shape. In 1997, Frauendorf
and Meng proposed that the total angular momentum vector of a rotating
triaxial nucleus may lie outside the three principal planes in the intrinsic
frame. Such an angular momentum geometry can, in the laboratory frame, give
rise to a pair of nearly degenerate $\Delta I = 1$ bands with
the same parity, i.e., chiral doublet bands~\cite{Frauendorf1997NPA}.
So far, more than 50 chiral candidates have been reported in odd-odd,
odd-$A$, and even-even nuclei that spread over $A \sim 80$~\cite{S.Y.Wang2011PLB},
100~\cite{Vaman2004PRL, Joshi2004PLB, Timar2004PLB, Nunez2004PRC, Y.X.Luo2009PLB,
Tonev2014PRL, Lieder2014PRL, Moon2018PLB}, 130~\cite{Starosta2001PRL, Koike2001PRC,
Hecht2001PRC, Hartley2001PRC, Zhu2003PRL, Koike2003PRC, Grodner2006PRL, S.Y.Wang2006PRC,
Mukhopadhyay2007PRL, Grodner2011PLB, Bujor2018PRC}, and 190 mass
regions~\cite{Balabanski2004PRC, Lawrie2008PRC}. For more details,
see reviews~\cite{J.Meng2010JPG, J.Meng2014IJMPE, Bark2014IJMPE, J.Meng2016PS,
Raduta2016PPNP, Frauendorf2018PS} and very recent data tables~\cite{B.W.Xiong2019ADNDT}.

During the process of investigating nuclear chirality, exploring novel chiral
phenomena and searching for new chiral candidates are the two fundamental
goals all the time. For the former, e.g., the multiple chiral doublets (M$\chi$D)
phenomenon, i.e., having multiple pairs of chiral doublet bands
in a single nucleus, was theoretically predicted and explored by the state-of-art
covariant density functional theory (CDFT)~\cite{J.Meng2006PRC, J.Peng2008PRC, J.M.Yao2009PRC,
J.Li2011PRC, J.Li2018PRC, B.Qi2018PRC, J.Peng2018PRC} and observed in
$^{133}$Ce~\cite{Ayangeakaa2013PRL}, $^{103}$Rh~\cite{Kuti2014PRL},
$^{78}$Br~\cite{C.Liu2016PRL}, $^{136}$Nd~\cite{Petrache2018PRC, Q.B.Chen2018PLB},
and $^{195}$Tl~\cite{Roy2018PLB}. These observations confirm the existence of
triaxial shapes coexistence~\cite{J.Meng2006PRC, Ayangeakaa2013PRL, Petrache2018PRC,
Roy2018PLB}, and reveal the stability of chiral geometry against the increasing of
intrinsic excitation energy~\cite{Droste2009EPJA, Q.B.Chen2010PRC, Hamamoto2013PRC,
Kuti2014PRL, H.Zhang2016CPC} and octupole correlations~\cite{C.Liu2016PRL}. For
the latter, the experimental evidence of chiral doublet bands was first observed in
the $A\sim 130$ mass region, and then followed by the $A\sim 100$, $190$, and
$80$ mass regions. These observations show that the nuclear chirality is not
a specific phenomenon that exists in only one nucleus or one mass region.

Both of the two fundamental goals and all of relevant observations mentioned
above encourage us to search for new candidates with chirality or M$\chi$D in new mass
regions. In Ref.~\cite{J.Peng2018PRC}, we explored the M$\chi$D in $A\sim 60$
mass region by the adiabatic and configuration-fixed constrained
CDFT for cobalt isotopes. It was found that there are high-$j$ particle(s) and
hole(s) configurations with prominent triaxially deformed shapes in these isotopes,
which suggests the possibility of chirality or multiple chirality in $A\sim 60$
mass region. However, the experimental energy spectra and electromagnetic
transition in these isotopes are rather rare at present.

We note that in Ref.~\cite{P.W.Zhao2011PLB}, the fully microscopic self-consistent
tilted axis cranking covariant density functional theory (TAC-CDFT) was applied to
investigate the observed dipole bands M1, M2, M3, and M4 in even-even nucleus
$^{60}$Ni~\cite{Torres2008PRC}. It was mentioned that bands M1 and M4 might be the possible
candidates for chiral doublet bands. However, due to the mean-field approximation,
the TAC can only give the description for the band M1. After that, there is neither
further theoretical nor experimental work to investigate bands M1 and M4 in $^{60}$Ni.
Therefore, whether the chirality exists in the bands M1 and M4 or not is still an
open problem.

The aim of the present work is to investigate the chirality in doublet bands M1 and M4
in $^{60}$Ni in a fully quantal model. As a quantal model coupling the collective
rotation and the single-particle motions, the particle rotor model (PRM) has been
widely used to describe the chiral doublet bands and achieved major successes~\cite{J.Meng2010JPG}.
In contrast to the TAC approach, PRM describes a system in the laboratory frame. The
total Hamiltonian is diagonalized with total angular momentum as a good quantum
number, and the energy splitting and quantum tunneling between the doublet
bands can be obtained directly. Moreover, the basic microscopic
inputs for PRM can be obtained from the constrained CDFT~\cite{J.Meng2006PRC,
J.Meng2016book, Ayangeakaa2013PRL, Lieder2014PRL, Kuti2014PRL, C.Liu2016PRL, Petrache2016PRC}.
Various versions of PRM have been developed to investigate the chiral
doublet bands with different kinds of configurations~\cite{Frauendorf1997NPA,
J.Peng2003PRC, Koike2004PRL, B.Qi2009PRC, H.Zhang2016CPC, Q.B.Chen2018PRC, Q.B.Chen2018PRC_v1,
Koike2003PRC, S.Q.Zhang2007PRC, S.Y.Wang2007PRC, S.Y.Wang2008PRC,
Lawrie2008PRC, Lawrie2010PLB, Shirinda2012EPJA, B.Qi2009PLB, B.Qi2011PRC,
Ayangeakaa2013PRL, B.Qi2013PRC, Lieder2014PRL, Kuti2014PRL,
Petrache2016PRC}. To describe the doublet bands M1 and M4 in $^{60}$Ni with
four quasi-particle configuration~\cite{Torres2008PRC, P.W.Zhao2011PLB}, a four-$j$
shell PRM is needed. Such version of PRM has already been
developed very recently and applied to describe the M$\chi$D in
$^{136}$Nd~\cite{Q.B.Chen2018PLB}.

In this letter, the four-$j$ shell PRM will be applied to study the energy
spectra and the electromagnetic transition probabilities of the doublet
bands M1 and M4 in $^{60}$Ni, and to explore the open problem on whether
or not the chirality exists in this doublet by examining their angular
momentum geometries.

%%%%%%%%%%%%%%%%%%%%%%%%%%%%%%%%%%%%%%%%%%%%%%%%%%%%%%%%%%
%                    begin  framework
%%%%%%%%%%%%%%%%%%%%%%%%%%%%%%%%%%%%%%%%%%%%%%%%%%%%%%%%%%

%\section{Theoretical framework}

The formalism of PRM with four-$j$ shell can be found in Ref.~\cite{Q.B.Chen2018PLB}.
The total wave function of PRM Hamiltonian is expanded into the strong
coupling basis
\begin{align}
 |IM\rangle=\sum_{K\phi} c_{K\phi}|IMK\phi\rangle,
\end{align}
with
\begin{align}
 |IMK\phi\rangle
  &=\frac{1}{\sqrt{2(1+\delta_{K0}\delta_{\phi,\bar{\phi}})}}\notag\\
  &\times \big(|IMK\rangle|\phi\rangle
  +(-1)^{I-K}|IM-K\rangle|\bar{\phi}\rangle\big),
\end{align}
where $|IMK\rangle$ is the Wigner function $\sqrt{\frac{2I+1}{8\pi^2}}D_{MK}^I$,
$|\phi\rangle$ is the product of the proton and neutron states those sitting
in the four-$j$ shells, and the $c_{K\phi}$ is the expansion coefficient
obtained by diagonalizing the PRM Hamiltonian. With the obtained wave functions,
the reduced transition probabilities $B(M1)$ and $B(E2)$ can be
calculated~\cite{Bohr1975}. In addition, one can also calculate the expectation
values of the angular momentum components~\cite{B.Qi2009PLB, Q.B.Chen2018PLB},
\begin{align}\label{eq1}
 J_k=\sqrt{\langle IM|\hat{J}_k^2|IM\rangle},
\end{align}
the probability distributions of the total angular momentum in the intrinsic
reference frame (azimuthal plot)~\cite{F.Q.Chen2017PRC, Q.B.Chen2018PRC_v1,
Streck2018PRC},
\begin{align}\label{eq2}
 \mathcal{P}(\theta,\varphi) &=2\pi \sum_{\phi^\prime}\Big|
 \sum_{K,\phi}c_{K,\phi}\sqrt{\frac{2I+1}{16\pi^2}}\big[
 D_{IK}^I(\psi, \theta,\pi-\varphi)\delta_{\phi^\prime, \phi}\notag\\
 &\quad +(-1)^{I-K}D_{I-K}^I(\psi, \theta,\pi-\varphi)\delta_{\phi^\prime, -\phi}
 \big]\Big|^2,
\end{align}
and the probability distributions of the total angular momentum components on
the three principle axes ($K$-plot)~\cite{B.Qi2009PLB, F.Q.Chen2017PRC},
\begin{align}\label{eq3}
 P_K=\sum_\phi |c_{K\phi}|^2.
\end{align}
Based on these analyses, one can study the angular momentum
geometries systematically to make an unambiguous judgment whether or
not the chiral geometry exists in the doublet bands.

%\section{Numerical details}

In the PRM calculations for the doublet bands M1 and M4 in $^{60}$Ni,
the configuration $\pi(1f_{7/2})^{-1}(2p_{3/2})^1\otimes \nu(1g_{9/2})^1
(1f_{5/2})^{-1}$~\cite{Torres2008PRC, P.W.Zhao2011PLB} is adopted. The deformation parameters
$\beta=0.27$ and $\gamma=19^\circ$ for this configuration at the bandhead were
obtained from the microscopic self-consistent TAC-CDFT calculations~\cite{P.W.Zhao2011PLB}.
With the rotation, the $\beta$ value decreases smoothly, while $\gamma$ value shows
a smoothly increasing tendency. In the PRM calculation, the deformation parameters
are fixed. To reproduce the energy spectra better, we use a deformation of $\beta=0.27$
and $\gamma=22^\circ$. The moment of inertia $\mathcal{J}_0 = 9.0~\hbar^2/\textrm{MeV}$
and Coriolis attenuation factor $\xi=0.98$ are adopted according to the experimental
energy spectra. For the electromagnetic transitions, the empirical intrinsic quadrupole moment
$Q_0=(3/\sqrt{5\pi})R_0^2 Z\beta$, and gyromagnetic ratios for rotor $g_R=Z/A$ and
for nucleons $g_{p(n)}=g_l+(g_s-g_l)/(2l+1)$ ($g_l=1(0)$ for protons (neutrons)
and $g_s=0.6g_{s}(\textrm{free})$)~\cite{Ring1980book} are adopted.

%%%%%%%%%%%%%%%%%%%%%%%%%%%%%%%%%%%%%%%%%%%%%%%%%%%%%%%%%%
%                    begin  results and discussion
%%%%%%%%%%%%%%%%%%%%%%%%%%%%%%%%%%%%%%%%%%%%%%%%%%%%%%%%%%

%\section{Results and discussion}

%\subsection{Energy spectra and $B(M1)/B(E2)$}

\begin{figure}[!ht]
  \begin{center}
    \includegraphics[width=5.0 cm, height=8.0 cm]{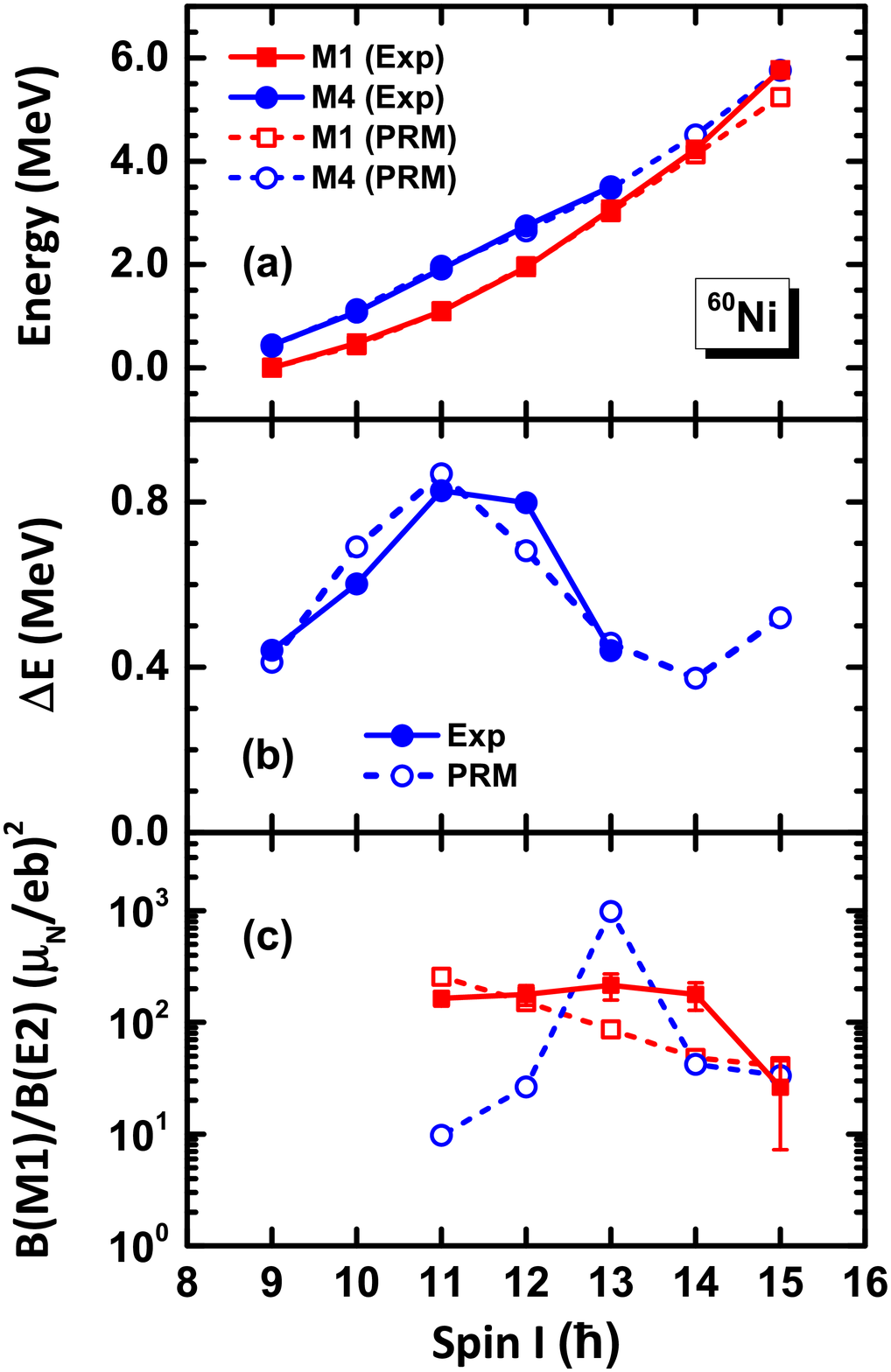} \quad
    \includegraphics[width=5.0 cm, height=8.0 cm]{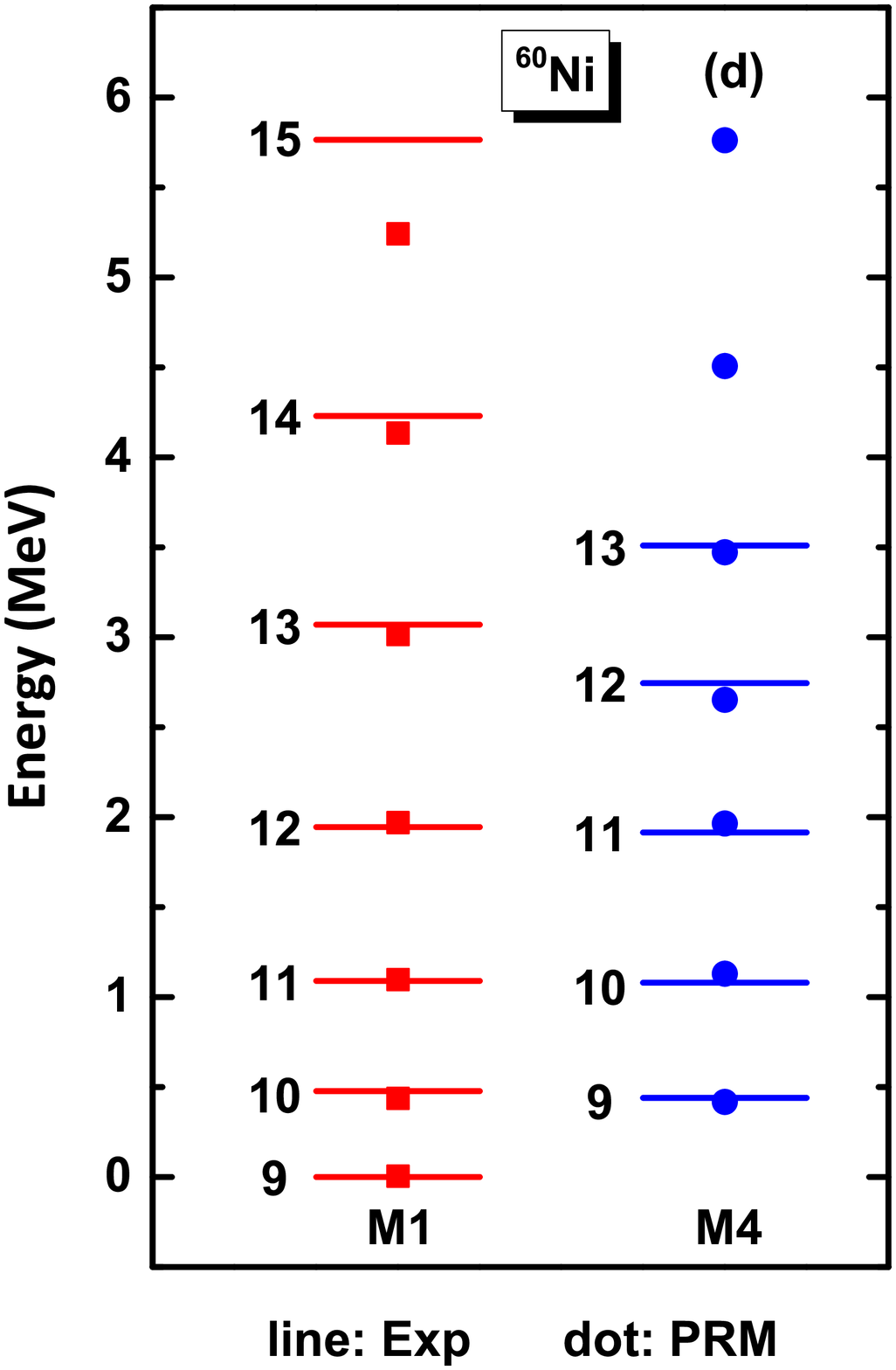}
    \caption{(a) Energy spectra as functions of spin for the bands M1 and M4 in
    $^{60}$Ni calculated by PRM in comparison with the data. (b) Theoretical and experimental
    energy difference between the doublet bands. (c) $B(M1)/B(E2)$ of bands M1 and M4
    calculated by PRM in comparison with the available data. (d) Energy level scheme
    of bands M1 and M4.}\label{fig1}
  \end{center}
\end{figure}

The calculated energy spectra for the bands M1 and M4 in $^{60}$Ni are
presented in Fig.~\ref{fig1}(a), together with the corresponding data.
The experimental energy spectra are reproduced excellently by the PRM
calculations. Such a good agreement can be more clearly seen by
showing the energy level scheme in Fig.~\ref{fig1}(d).

Being a fully quantal model, PRM is capable of reproducing the energy splitting $\Delta E$
between the doublet bands for the whole observed spin region. This is illustrated in
Fig.~\ref{fig1}(b). The $\Delta E$ increases firstly from $I=9$ to $11\hbar$, and then
decreases up to $I=14\hbar$. At $I=15\hbar$, an increasing trend is observed once again in
the PRM results. It is known that in an ideal chiral system with the particle-hole
configuration $\pi(1h_{11/2})^1 \otimes \nu(1h_{11/2})^{-1}$ and a rotor with the
deformation parameter $\gamma=30^\circ$, the $\Delta E$ is small (less than 400 keV)
and shows a trend that decreases firstly and then increases~\cite{Frauendorf1997NPA,
Q.B.Chen2018PRC_v1}. Here, at $I\geq 12\hbar$, the $\Delta E$ shows the similar variation
trend, giving a hint that chirality might exist in bands M1 and M4. The large
$\Delta E$ (higher than 400 keV) could be owed to the small triaxial
deformation~\cite{B.Qi2009PRC, H.Zhang2016CPC}. Therefore, it would be very
interesting to extend the spectrum of band M4, which only reaches to $I=13\hbar$
currently, to higher spins to further verify the theoretical calculations.

In Fig.~\ref{fig1}(c), the $B(M1)/B(E2)$ values of bands M1 and M4
calculated by PRM in comparison with the available data
are shown. One observes that the PRM calculations show a good
agreement with the data at $I=11$ and $12\hbar$
for band M1. At $I=13$ and $14\hbar$, the calculated
$B(M1)/B(E2)$ is smaller than the data. This is because the decrease
of $\beta$ with the rotation~\cite{P.W.Zhao2011PLB} is not taken
into account in the present PRM calculations. The calculated $B(M1)/B(E2)$
for band M4 at $I=13\hbar$ is very large. It is caused by, as shown later,
the wave function structure changes dramatically from $I=12$ to $13\hbar$.
The calculated $B(M1)/B(E2)$ values of bands M1 and M4 at $I=14$ and
$15\hbar$ are similar, which further implies the chirality exists.
Therefore, further experimental efforts on extracting electromagnetic
transition data for band M4 are highly demanded to obtain solid evidence.

%\subsection{Angular momenta}

\begin{figure}[!ht]
  \begin{center}
    \includegraphics[width=10.5 cm]{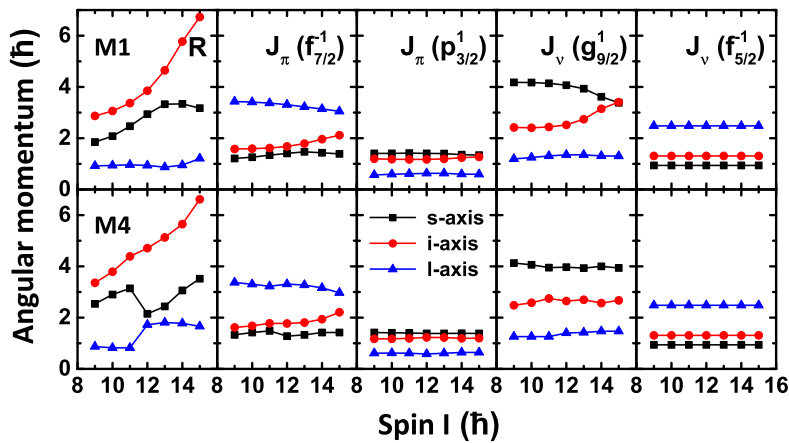}
    \caption{The root mean square components along the short ($s$-, squares),
    intermediate ($i$-, circles), and long ($l$-, triangles) axes of the rotor,
    valence protons, and valence neutrons angular momenta calculated as functions
    of spin by PRM for the doublet bands M1 and M4 in $^{60}$Ni.}\label{fig2}
  \end{center}
\end{figure}

The rotational motion of triaxial nuclei attains a chiral character if the
angular momentum has substantial projections on all three principal
axes of the triaxially deformed nucleus~\cite{Frauendorf1997NPA}.
The successes in reproducing the energy spectra and available electromagnetic
transition probabilities for the doublet bands M1 and M4 in
$^{60}$Ni motivate us to investigate the expectation values of the
squared angular momentum components along the short ($s$-), intermediate
($i$-), and long ($l$-) axes for the rotor, valence protons, and valence
neutrons. As shown in Fig.~\ref{fig2}, the substantial projections of angular momentum
on three principal axes can be observed for bands M1 and M4.

For both bands M1 and M4, the collective core angular momentum mainly
aligns along the $i$-axis in the whole spin region, because it has the
largest moment of inertia. In band M4, the $s$- and $l$- components of
the rotor angular momentum exhibit discontinuous behaviors
from $I=11$ to $12\hbar$. This is understood as the reason of abrupt
increases of $B(M1)/B(E2)$ values, as discussed previously. The angular
momenta of the $f_{7/2}$ valence proton and $f_{5/2}$ valence neutron holes
mainly align along the $l$-axis, and that of valence neutron $g_{9/2}$ particle
mainly along the $s$-axis. At high spin region in band M1, the $i$-component
($s$-component) of $g_{9/2}$ particle gradually increases (decreases). For $p_{3/2}$ proton,
the $s$- and $i$- components are similar to each other. Such orientations form
the chiral geometry of aplanar rotation.

%\subsection{Azimuthal plots}

\begin{figure}[!ht]
  \begin{center}
    \includegraphics[width=10.6 cm]{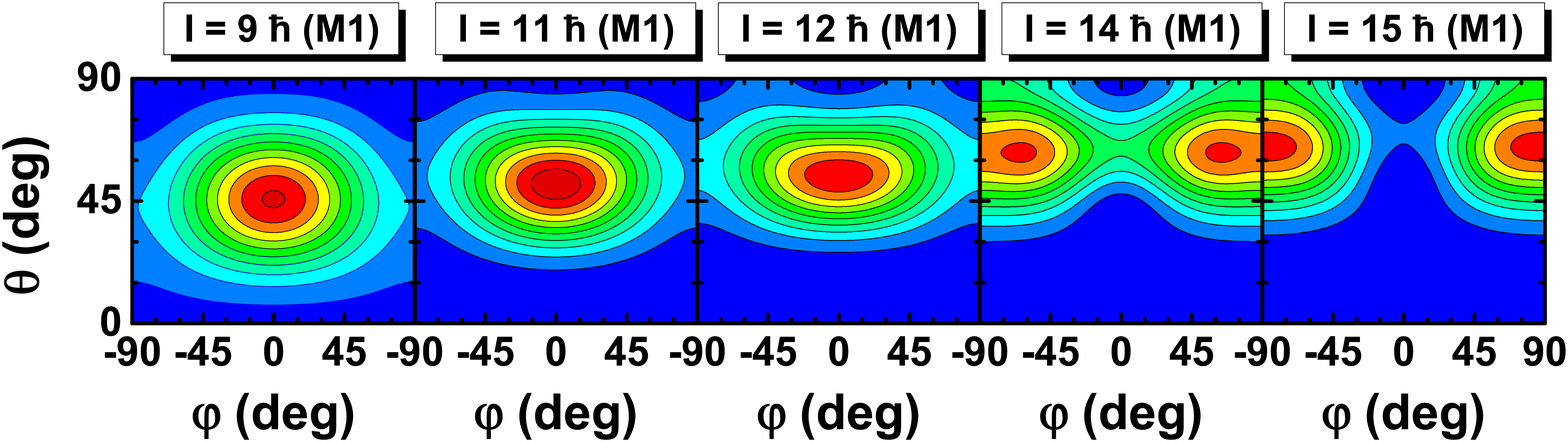}\\
    ~~\\
    \includegraphics[width=10.6 cm]{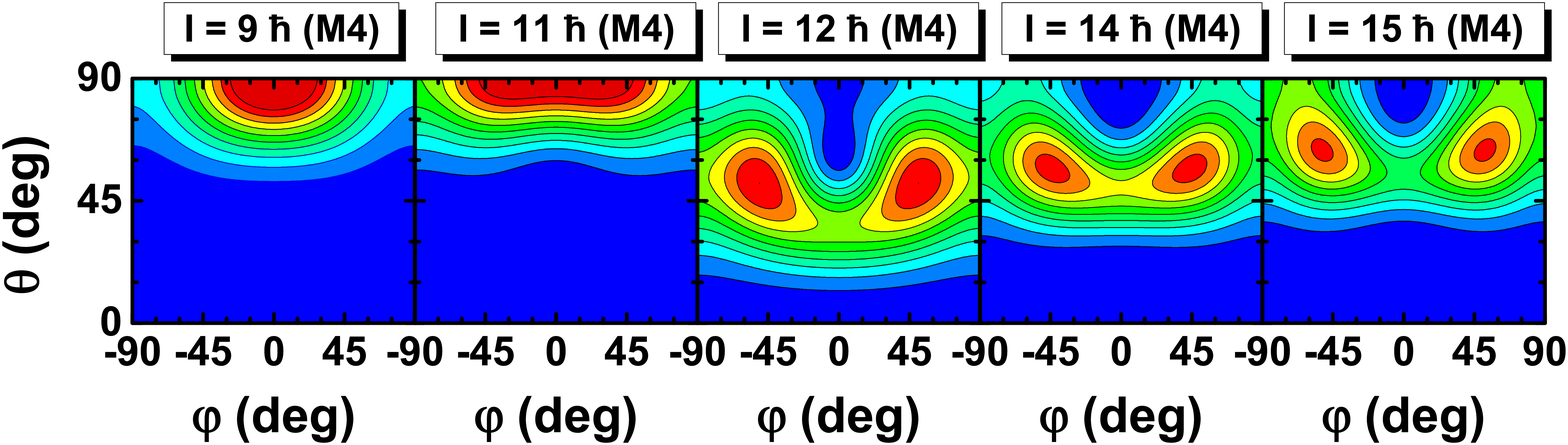}
    \caption{The azimuthal plots, i.e., profiles for the orientation of the angular momentum
    on the $(\theta, \varphi)$ plane calculated by PRM at $I=9$, $11$, $12$, $14$, and
    $15\hbar$ for the doublet bands M1 and M4 in $^{60}$Ni.}\label{fig4}
  \end{center}
\end{figure}

The spontaneous chiral symmetry breaking or chiral geometry is realized by the
total angular momentum lying outside the three principal planes in the intrinsic
frame~\cite{Frauendorf1997NPA}. In order to visualize the angular momentum geometry
in the intrinsic frame, the azimuthal plots~\cite{F.Q.Chen2017PRC, Q.B.Chen2018PRC_v1,
Streck2018PRC}, i.e., profiles $\mathcal{P}(\theta,\varphi)$
for the orientation of the angular momentum on the $(\theta, \varphi)$ plane
calculated by PRM using Eq.~(\ref{eq2}) are shown in Fig.~\ref{fig4} for
the doublet bands M1 and M4 in $^{60}$Ni at $I=9$, $11$, $12$,
$14$, and $15\hbar$. We emphasize that $\theta$ is the
angle between the total spin $\bm{I}$ and the $l$-axis, and $\varphi$ is
the angle between the projection of $\bm{I}$ onto the $si$-plane and the $s$-axis.
As is shown in Fig.~\ref{fig4}, the azimuthal plots are symmetric with
respect to $\varphi=0^\circ$. This is because the broken chiral
symmetry in the intrinsic frame has been restored in PRM.

For $I=9$ (bandhead) and $11\hbar$ (kink of $\Delta E$), the profiles
for the orientation of the angular momentum for band M1 have only one single
peak at $(\theta \sim 45^\circ, \varphi=0^\circ)$, which suggests
that the angular momentum stays within the $sl$-plane. Instead, the
profiles for band M4 peak at $(\theta=90^\circ, \varphi=0^\circ)$,
corresponding to a principal axis rotation with respect to $s$-axis. Such two orientations
do not form a chiral geometry. Therefore, not chirality is shown for $I=9$-$11\hbar$.
This is also the reason why the energy splitting $\Delta E$ increases at this spin
region, as shown in Fig.~\ref{fig1}(b).

For $I=12\hbar$, the profile for band M1 still peaks at $\varphi=0^\circ$.
However, that for band M4 shows a node around $(\theta\sim 60^\circ, \varphi=0^\circ)$
with the onset of two peaks locating at $(\theta\sim 45^\circ, \varphi \sim 45^\circ)$
and $(\theta\sim 45^\circ, \varphi \sim -45^\circ)$, respectively. The appearances of
node and two peaks are consistent with the picture of a 0-phonon state in band M1 and
1-phonon vibration in band M4~\cite{F.Q.Chen2017PRC, Q.B.Chen2018PRC_v1, Streck2018PRC}.
Therefore, the chiral vibration is demonstrated for $I=12\hbar$. Due to the chiral
vibration, the $i$- ($s$-) component of the
rotator angular momentum at $I=12\hbar$ for band M4 is larger (smaller) than
that for band M1, as shown in Fig.~\ref{fig2}.

For $I=14\hbar$, two peaks corresponding to aplanar orientations are found in the
both of doublet bands, i.e., $(\theta \sim 60^\circ, \varphi \sim 60^\circ)$ and $(\theta
\sim 60^\circ, \varphi \sim -60^\circ)$ for band M1, while $(\theta \sim 55^\circ,
\varphi \sim 50^\circ)$ and $(\theta \sim 55^\circ, \varphi \sim -50^\circ)$ for band M4.
These features could be understood as a realization of static chirality,
and hence give the lowest $\Delta E$ as shown in Fig.~\ref{fig1}(b).

For $I=15\hbar$, the peaks for band M1 move toward to $(\theta \sim 60^\circ,
\varphi \sim 90^\circ)$ and $(\theta \sim 60^\circ, \varphi \sim -90^\circ)$,
namely in the $il$-plane and close to $i$-axis. This is mainly driven by the
gradual increasing of $i$-components of the rotor, and valence neutron $g_{9/2}$
particle, and valence proton $f_{7/2}$ hole angular momenta, as presented
in Fig.~\ref{fig2}. The peaks for the azimuthal plot for band M4 locate
at $(\theta\sim 60^\circ, \varphi \sim 60^\circ)$ and $(\theta\sim 60^\circ,
\varphi \sim -60^\circ)$. At this spin, bands M1 and M4 attain vibration
character again, which is now with respect to $il$-plane. As a consequence,
their energy difference $\Delta E$, as shown in Fig.~\ref{fig1}(b),
increases~\cite{Q.B.Chen2013PRC, Q.B.Chen2016PRC}.

%\subsection{$K$-plots}

\begin{figure*}[!ht]
  \begin{center}
    \includegraphics[width=5.3 cm]{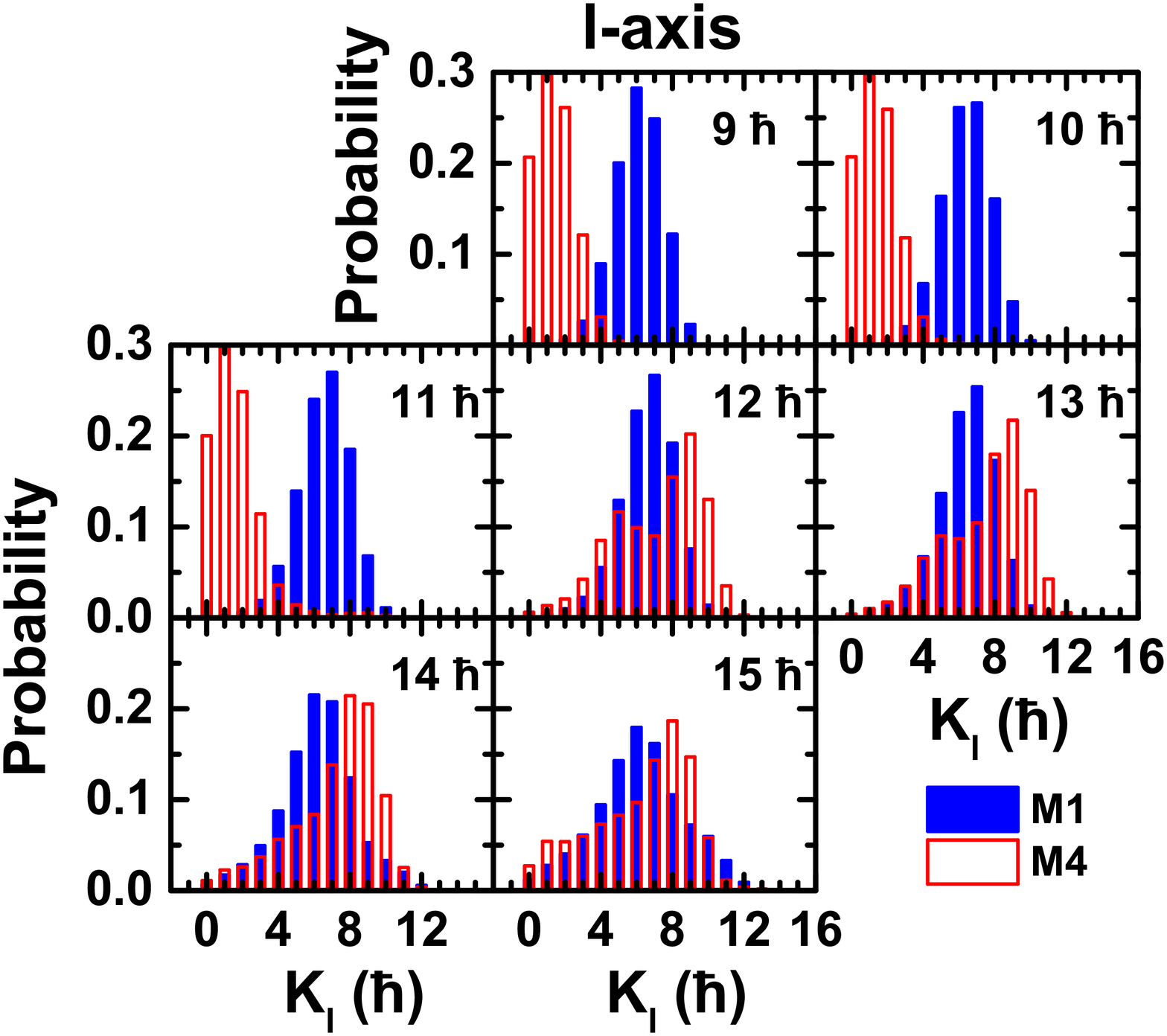}~~
    \includegraphics[width=5.3 cm]{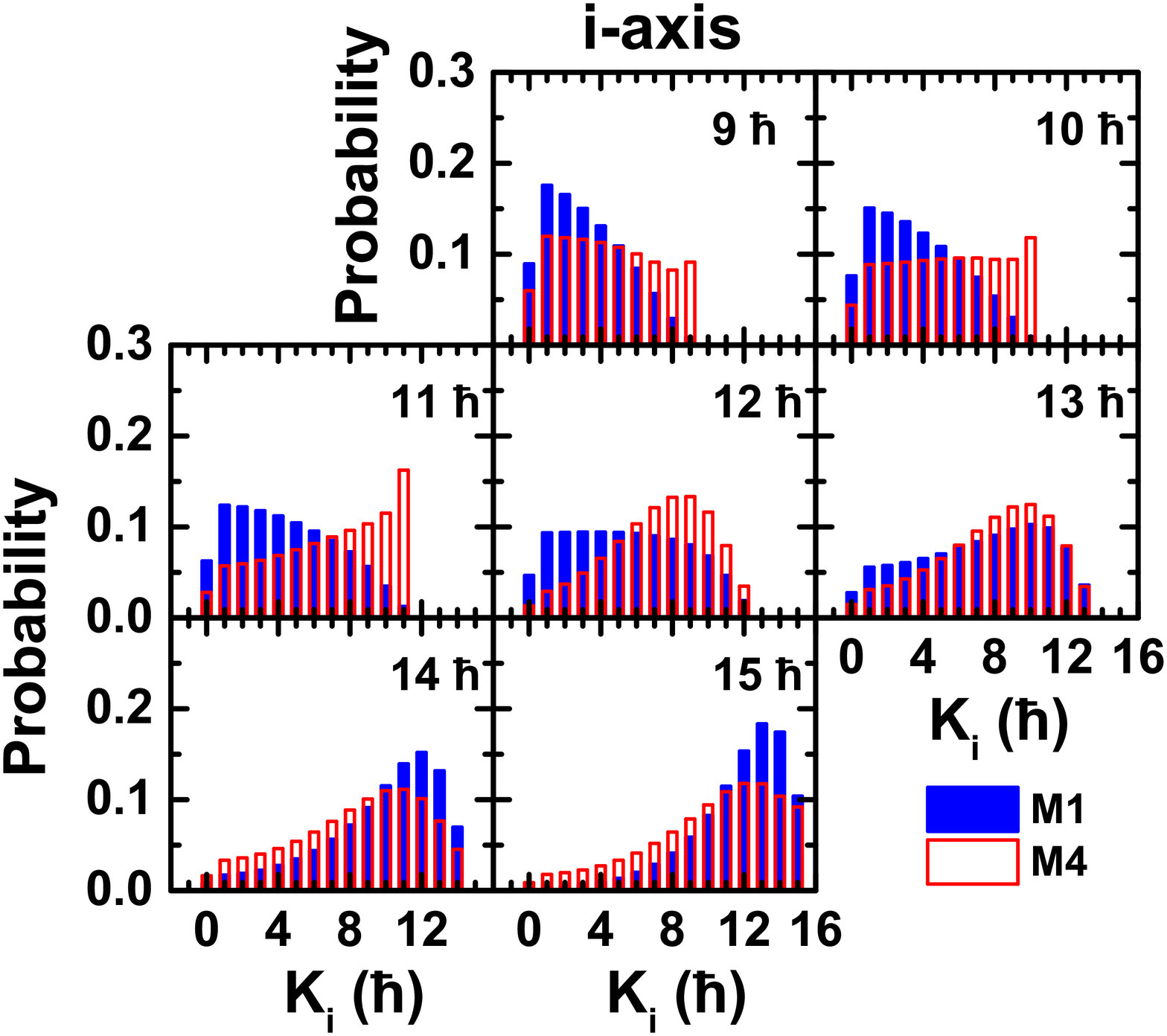}~~
    \includegraphics[width=5.3 cm]{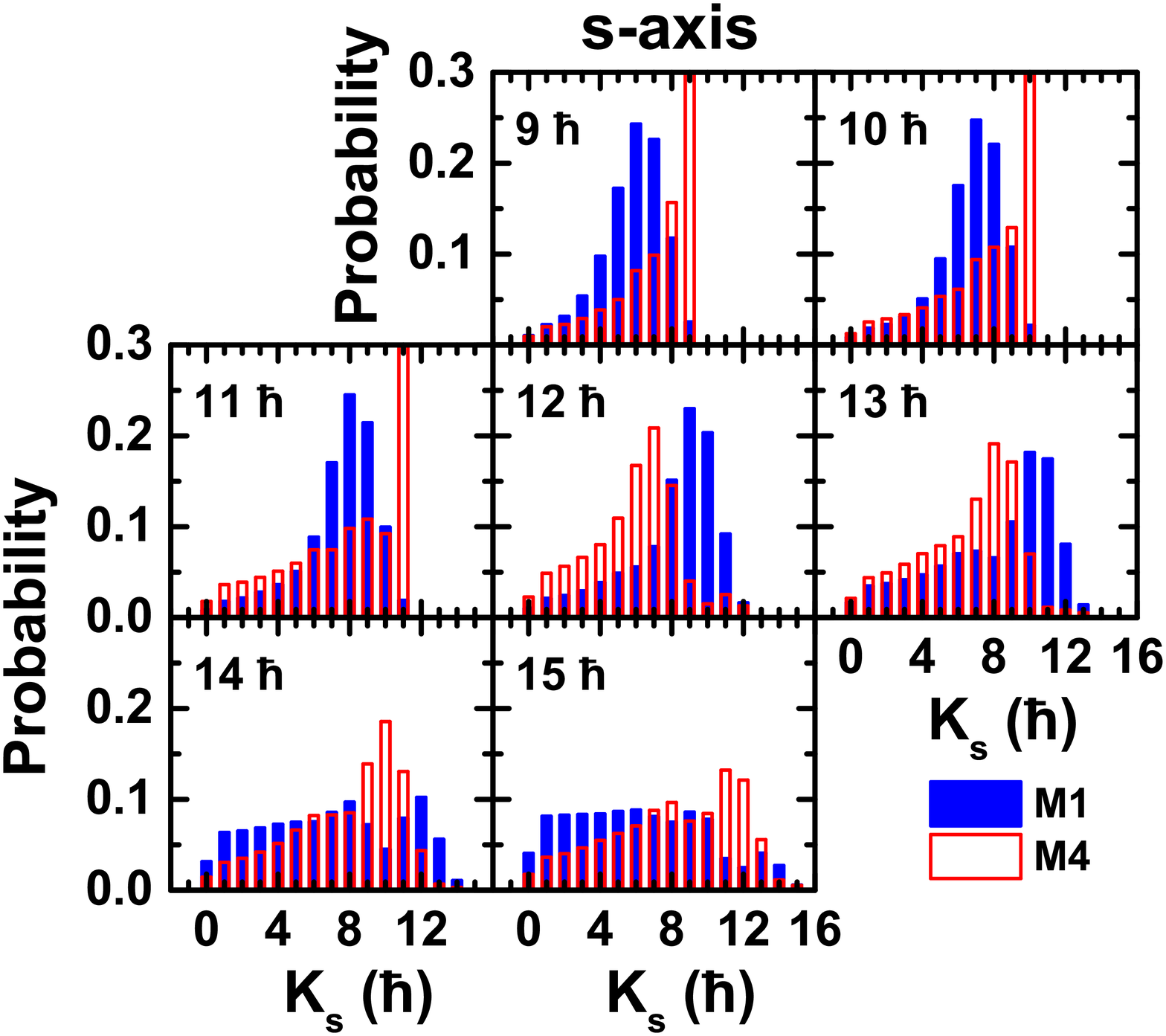}
    \caption{The $K$-plots, i.e., $K$-distributions for the angular momentum
    on the long ($l$-), intermediate ($i$), and short ($s$-) axes calculated by PRM
    for the doublet bands M1 and M4 in $^{60}$Ni.}\label{fig3}
  \end{center}
\end{figure*}

To further understand the evolution of the chirality with spin, in Fig.~\ref{fig3},
the $K$-plots, i.e., $K$-distributions for the angular momentum on the $l$-, $i$-,
and $s$- axes calculated by PRM for the doublet bands M1 and M4 in $^{60}$Ni are
displayed. As seen in the figures, the evolutions of the rotational modes from
no chirality at $I=9$-$11\hbar$, to the chiral vibration at $I=12\hbar$, then to
the static chirality at $I=13$-$14\hbar$, and finally to the second vibration
at $I=15\hbar$ are exhibited clearly.

For $I=9$-$11\hbar$, the peaks of $K_l$ locate around $K_l=6\hbar$ for band M1, while
at $K_l=1\hbar$ for band M4. The $K_i$-distribution peaks at $K_i=1\hbar$ for band M1,
while spreads widely for band M4. The $K_s$-distributions of both bands peak at large
$K_s$ value. All of these are in accordance with the features observed in the
azimuthal plots shown in Fig.~\ref{fig4}. Namely, the angular momentum of band M1
stays within the $sl$-plane, while that of band M4 aligns along $s$-axis.

For $I=12\hbar$, $K_l$ distribution of band M4 shows a rapid change, which
is caused by the discontinuous variation of the $l$-component of rotor angular
momentum, as shown in Fig.~\ref{fig2}. The $K_i$-distribution spreads around $K_i = 0$
for band M1, whereas it almost vanishes for band M4. This is in accordance with the
interpretation of the chiral vibration with respect to the $sl$-plane where the
zero-phonon state (band M1) is symmetric with respect to $K_i = 0$ and the
one-phonon state (band M4) is antisymmetric.

For $I=13$ and $14\hbar$, the $K_i$-distributions of bands M1 and M4 are rather
similar. The position of peak of $K_l$- ($K_s$) distribution for band
M1 is a bit smaller (larger) than those for band M4. Such differences lead
the different azimuthal plots shown in Fig.~\ref{fig4} and lead that the
energy splitting of bands M1 and M4 is a bit large ($\sim$ 400 keV),
though they are, in fact, in the static chirality region.

For $I=15\hbar$, the $K_l$- and $K_i$- distributions for bands M1 and M4 are similar.
However, for the $K_s$-distribution, they are different. The most probable value
spreads at $K_s\sim 1\hbar$ for band M1, while appears at $K_s \sim 11\hbar$ for
band M4. This further supports the appearance of second chiral vibration with
respect to $il$-plane.

%%%%%%%%%%%%%%%%%%%%%%%%%%%%%%%%%%%%%%%%%%%%%%%%%%%%%%%%
%                    begin  summary
%%%%%%%%%%%%%%%%%%%%%%%%%%%%%%%%%%%%%%%%%%%%%%%%%%%%%%%%

%\section{Summary}

In summary, the open problem on whether or not the chirality exists in doublet
bands M1 and M4 in light-mass even-even nucleus $^{60}$Ni is studied by adopting
the recently developed fully quantal four-$j$ shells triaxial particle
rotor model. The corresponding experimental energy spectra, energy differences
between the doublet bands, and the available $B(M1)/B(E2)$ values are successfully
reproduced. The analyses based on the angular momentum components, the
azimuthal plots, and the $K$-plots suggest that the chiral modes exist at
$I\geq 12\hbar$. Namely, there is no indication of chirality at $I\leq 11\hbar$.
A chiral vibration appears at $I=12\hbar$, then changes
to nearly static chirality at $I=14\hbar$, and finally evolves
to another type of chiral vibration at $I=15\hbar$.

Further experimental efforts on extending the level scheme and extracting
electromagnetic transition data for band M4 are highly demanded to obtain
solid evidence. According to current investigation, we would also like to
attract more experimental and theoretical efforts on the investigation of
chirality or multiple chirality in the $A\sim 60$ mass region and even in
the lighter-mass region.

\section*{Acknowledgements}

The authors thank Professor Jie Meng for helpful discussions.
This work is supported by the National Natural Science
Foundation of China (NSFC) under Grants No. 11775026 and 11875027,
and the Deutsche Forschungsgemeinschaft (DFG) and NSFC
through funds provided to the Sino-German CRC110 ``Symmetries and
the Emergence of Structure in QCD'' (DFG Grant No. TRR110 and NSFC
Grant No. 11621131001).

%%%%%%%%%%%%%%%%%%%%%%%%%%%%%%%%%%%%%%%%%%%%%%%%%%%%%%%%
%                  begin refereee
%%%%%%%%%%%%%%%%%%%%%%%%%%%%%%%%%%%%%%%%%%%%%%%%%%%%%%%%

\end{CJK}


\begin{thebibliography}{70}
\expandafter\ifx\csname natexlab\endcsname\relax\def\natexlab#1{#1}\fi
\expandafter\ifx\csname bibnamefont\endcsname\relax
  \def\bibnamefont#1{#1}\fi
\expandafter\ifx\csname bibfnamefont\endcsname\relax
  \def\bibfnamefont#1{#1}\fi
\expandafter\ifx\csname citenamefont\endcsname\relax
  \def\citenamefont#1{#1}\fi
\expandafter\ifx\csname url\endcsname\relax
  \def\url#1{\texttt{#1}}\fi
\expandafter\ifx\csname urlprefix\endcsname\relax\def\urlprefix{URL }\fi
\providecommand{\bibinfo}[2]{#2}
\providecommand{\eprint}[2][]{\url{#2}}

\bibitem[{\citenamefont{Frauendorf and Meng}(1997)}]{Frauendorf1997NPA}
\bibinfo{author}{\bibfnamefont{S.}~\bibnamefont{Frauendorf}} \bibnamefont{and}
  \bibinfo{author}{\bibfnamefont{J.}~\bibnamefont{Meng}},
  \bibinfo{journal}{Nucl. Phys. A} \textbf{\bibinfo{volume}{617}},
  \bibinfo{pages}{131} (\bibinfo{year}{1997}).

\bibitem[{\citenamefont{Wang et~al.}(2011)\citenamefont{Wang, Qi, Liu, Zhang,
  Hua, Li, Chen, Zhu, Meng, Wyngaardt et~al.}}]{S.Y.Wang2011PLB}
\bibinfo{author}{\bibfnamefont{S.~Y.} \bibnamefont{Wang}},
  \bibinfo{author}{\bibfnamefont{B.}~\bibnamefont{Qi}},
  \bibinfo{author}{\bibfnamefont{L.}~\bibnamefont{Liu}},
  \bibinfo{author}{\bibfnamefont{S.~Q.} \bibnamefont{Zhang}},
  \bibinfo{author}{\bibfnamefont{H.}~\bibnamefont{Hua}},
  \bibinfo{author}{\bibfnamefont{X.~Q.} \bibnamefont{Li}},
  \bibinfo{author}{\bibfnamefont{Y.~Y.} \bibnamefont{Chen}},
  \bibinfo{author}{\bibfnamefont{L.~H.} \bibnamefont{Zhu}},
  \bibinfo{author}{\bibfnamefont{J.}~\bibnamefont{Meng}},
  \bibinfo{author}{\bibfnamefont{S.~M.} \bibnamefont{Wyngaardt}},
  \bibnamefont{et~al.}, \bibinfo{journal}{Phys. Lett. B}
  \textbf{\bibinfo{volume}{703}}, \bibinfo{pages}{40} (\bibinfo{year}{2011}).

\bibitem[{\citenamefont{Vaman et~al.}(2004)\citenamefont{Vaman, Fossan, Koike,
  Starosta, Lee, and Macchiavelli}}]{Vaman2004PRL}
\bibinfo{author}{\bibfnamefont{C.}~\bibnamefont{Vaman}},
  \bibinfo{author}{\bibfnamefont{D.~B.} \bibnamefont{Fossan}},
  \bibinfo{author}{\bibfnamefont{T.}~\bibnamefont{Koike}},
  \bibinfo{author}{\bibfnamefont{K.}~\bibnamefont{Starosta}},
  \bibinfo{author}{\bibfnamefont{I.~Y.} \bibnamefont{Lee}}, \bibnamefont{and}
  \bibinfo{author}{\bibfnamefont{A.~O.} \bibnamefont{Macchiavelli}},
  \bibinfo{journal}{Phys. Rev. Lett.} \textbf{\bibinfo{volume}{92}},
  \bibinfo{pages}{032501} (\bibinfo{year}{2004}).

\bibitem[{\citenamefont{Joshi et~al.}(2004)\citenamefont{Joshi, Jenkins,
  Raddon, Simons, Wadsworth, Wilkinson, Fossan, Koike, Starosta, Vaman
  et~al.}}]{Joshi2004PLB}
\bibinfo{author}{\bibfnamefont{P.}~\bibnamefont{Joshi}},
  \bibinfo{author}{\bibfnamefont{D.~G.} \bibnamefont{Jenkins}},
  \bibinfo{author}{\bibfnamefont{P.~M.} \bibnamefont{Raddon}},
  \bibinfo{author}{\bibfnamefont{A.~J.} \bibnamefont{Simons}},
  \bibinfo{author}{\bibfnamefont{R.}~\bibnamefont{Wadsworth}},
  \bibinfo{author}{\bibfnamefont{A.~R.} \bibnamefont{Wilkinson}},
  \bibinfo{author}{\bibfnamefont{D.~B.} \bibnamefont{Fossan}},
  \bibinfo{author}{\bibfnamefont{T.}~\bibnamefont{Koike}},
  \bibinfo{author}{\bibfnamefont{K.}~\bibnamefont{Starosta}},
  \bibinfo{author}{\bibfnamefont{C.}~\bibnamefont{Vaman}},
  \bibnamefont{et~al.}, \bibinfo{journal}{Phys. Lett. B}
  \textbf{\bibinfo{volume}{595}}, \bibinfo{pages}{135} (\bibinfo{year}{2004}).

\bibitem[{\citenamefont{Tim\'{a}r et~al.}(2004)\citenamefont{Tim\'{a}r, Joshi,
  Starosta, Dimitrov, Fossan, Molnar, Sohler, Wadsworth, Algora, Bednarczyk
  et~al.}}]{Timar2004PLB}
\bibinfo{author}{\bibfnamefont{J.}~\bibnamefont{Tim\'{a}r}},
  \bibinfo{author}{\bibfnamefont{P.}~\bibnamefont{Joshi}},
  \bibinfo{author}{\bibfnamefont{K.}~\bibnamefont{Starosta}},
  \bibinfo{author}{\bibfnamefont{V.~I.} \bibnamefont{Dimitrov}},
  \bibinfo{author}{\bibfnamefont{D.~B.} \bibnamefont{Fossan}},
  \bibinfo{author}{\bibfnamefont{J.}~\bibnamefont{Molnar}},
  \bibinfo{author}{\bibfnamefont{D.}~\bibnamefont{Sohler}},
  \bibinfo{author}{\bibfnamefont{R.}~\bibnamefont{Wadsworth}},
  \bibinfo{author}{\bibfnamefont{A.}~\bibnamefont{Algora}},
  \bibinfo{author}{\bibfnamefont{P.}~\bibnamefont{Bednarczyk}},
  \bibnamefont{et~al.}, \bibinfo{journal}{Phys. Lett. B}
  \textbf{\bibinfo{volume}{598}}, \bibinfo{pages}{178} (\bibinfo{year}{2004}).

\bibitem[{\citenamefont{Alc\'antara-N\'u\~nez
  et~al.}(2004)\citenamefont{Alc\'antara-N\'u\~nez, Oliveira, Cybulska, Medina,
  Rao, Ribas, Rizzutto, Seale, Falla-Sotelo, Wiedemann et~al.}}]{Nunez2004PRC}
\bibinfo{author}{\bibfnamefont{J.~A.} \bibnamefont{Alc\'antara-N\'u\~nez}},
  \bibinfo{author}{\bibfnamefont{J.~R.~B.} \bibnamefont{Oliveira}},
  \bibinfo{author}{\bibfnamefont{E.~W.} \bibnamefont{Cybulska}},
  \bibinfo{author}{\bibfnamefont{N.~H.} \bibnamefont{Medina}},
  \bibinfo{author}{\bibfnamefont{M.~N.} \bibnamefont{Rao}},
  \bibinfo{author}{\bibfnamefont{R.~V.} \bibnamefont{Ribas}},
  \bibinfo{author}{\bibfnamefont{M.~A.} \bibnamefont{Rizzutto}},
  \bibinfo{author}{\bibfnamefont{W.~A.} \bibnamefont{Seale}},
  \bibinfo{author}{\bibfnamefont{F.}~\bibnamefont{Falla-Sotelo}},
  \bibinfo{author}{\bibfnamefont{K.~T.} \bibnamefont{Wiedemann}},
  \bibnamefont{et~al.}, \bibinfo{journal}{Phys. Rev. C}
  \textbf{\bibinfo{volume}{69}}, \bibinfo{pages}{024317}
  (\bibinfo{year}{2004}).

\bibitem[{\citenamefont{Luo et~al.}(2009)\citenamefont{Luo, Zhu, Hamilton,
  Rasmussen, Ramayya, Goodin, Li, Hwang, Almehed, Frauendorf
  et~al.}}]{Y.X.Luo2009PLB}
\bibinfo{author}{\bibfnamefont{Y.~X.} \bibnamefont{Luo}},
  \bibinfo{author}{\bibfnamefont{S.~J.} \bibnamefont{Zhu}},
  \bibinfo{author}{\bibfnamefont{J.~H.} \bibnamefont{Hamilton}},
  \bibinfo{author}{\bibfnamefont{J.~O.} \bibnamefont{Rasmussen}},
  \bibinfo{author}{\bibfnamefont{A.~V.} \bibnamefont{Ramayya}},
  \bibinfo{author}{\bibfnamefont{C.}~\bibnamefont{Goodin}},
  \bibinfo{author}{\bibfnamefont{K.}~\bibnamefont{Li}},
  \bibinfo{author}{\bibfnamefont{J.~K.} \bibnamefont{Hwang}},
  \bibinfo{author}{\bibfnamefont{D.}~\bibnamefont{Almehed}},
  \bibinfo{author}{\bibfnamefont{S.}~\bibnamefont{Frauendorf}},
  \bibnamefont{et~al.}, \bibinfo{journal}{Phys. Lett. B}
  \textbf{\bibinfo{volume}{670}}, \bibinfo{pages}{307} (\bibinfo{year}{2009}).

\bibitem[{\citenamefont{Tonev et~al.}(2014)\citenamefont{Tonev, Yavahchova,
  Goutev, de~Angelis, Petkov, Bhowmik, Singh, Muralithar, Madhavan, Kumar
  et~al.}}]{Tonev2014PRL}
\bibinfo{author}{\bibfnamefont{D.}~\bibnamefont{Tonev}},
  \bibinfo{author}{\bibfnamefont{M.~S.} \bibnamefont{Yavahchova}},
  \bibinfo{author}{\bibfnamefont{N.}~\bibnamefont{Goutev}},
  \bibinfo{author}{\bibfnamefont{G.}~\bibnamefont{de~Angelis}},
  \bibinfo{author}{\bibfnamefont{P.}~\bibnamefont{Petkov}},
  \bibinfo{author}{\bibfnamefont{R.~K.} \bibnamefont{Bhowmik}},
  \bibinfo{author}{\bibfnamefont{R.~P.} \bibnamefont{Singh}},
  \bibinfo{author}{\bibfnamefont{S.}~\bibnamefont{Muralithar}},
  \bibinfo{author}{\bibfnamefont{N.}~\bibnamefont{Madhavan}},
  \bibinfo{author}{\bibfnamefont{R.}~\bibnamefont{Kumar}},
  \bibnamefont{et~al.}, \bibinfo{journal}{Phys. Rev. Lett.}
  \textbf{\bibinfo{volume}{112}}, \bibinfo{pages}{052501}
  (\bibinfo{year}{2014}).

\bibitem[{\citenamefont{Lieder et~al.}(2014)\citenamefont{Lieder, Lieder, Bark,
  Chen, Zhang, Meng, Lawrie, Lawrie, Bvumbi, Kheswa et~al.}}]{Lieder2014PRL}
\bibinfo{author}{\bibfnamefont{E.~O.} \bibnamefont{Lieder}},
  \bibinfo{author}{\bibfnamefont{R.~M.} \bibnamefont{Lieder}},
  \bibinfo{author}{\bibfnamefont{R.~A.} \bibnamefont{Bark}},
  \bibinfo{author}{\bibfnamefont{Q.~B.} \bibnamefont{Chen}},
  \bibinfo{author}{\bibfnamefont{S.~Q.} \bibnamefont{Zhang}},
  \bibinfo{author}{\bibfnamefont{J.}~\bibnamefont{Meng}},
  \bibinfo{author}{\bibfnamefont{E.~A.} \bibnamefont{Lawrie}},
  \bibinfo{author}{\bibfnamefont{J.~J.} \bibnamefont{Lawrie}},
  \bibinfo{author}{\bibfnamefont{S.~P.} \bibnamefont{Bvumbi}},
  \bibinfo{author}{\bibfnamefont{N.~Y.} \bibnamefont{Kheswa}},
  \bibnamefont{et~al.}, \bibinfo{journal}{Phys. Rev. Lett.}
  \textbf{\bibinfo{volume}{112}}, \bibinfo{pages}{202502}
  (\bibinfo{year}{2014}).

\bibitem[{\citenamefont{Moon et~al.}(2018)\citenamefont{Moon, Moon, Dracoulis,
  Bark, Byrne, Davidson, Lane, Kib\'{e}di, Wilson, Yuan et~al.}}]{Moon2018PLB}
\bibinfo{author}{\bibfnamefont{B.}~\bibnamefont{Moon}},
  \bibinfo{author}{\bibfnamefont{C.-B.} \bibnamefont{Moon}},
  \bibinfo{author}{\bibfnamefont{G.}~\bibnamefont{Dracoulis}},
  \bibinfo{author}{\bibfnamefont{R.}~\bibnamefont{Bark}},
  \bibinfo{author}{\bibfnamefont{A.}~\bibnamefont{Byrne}},
  \bibinfo{author}{\bibfnamefont{P.}~\bibnamefont{Davidson}},
  \bibinfo{author}{\bibfnamefont{G.}~\bibnamefont{Lane}},
  \bibinfo{author}{\bibfnamefont{T.}~\bibnamefont{Kib\'{e}di}},
  \bibinfo{author}{\bibfnamefont{A.}~\bibnamefont{Wilson}},
  \bibinfo{author}{\bibfnamefont{C.}~\bibnamefont{Yuan}}, \bibnamefont{et~al.},
  \bibinfo{journal}{Phys. Lett. B} \textbf{\bibinfo{volume}{782}},
  \bibinfo{pages}{602 } (\bibinfo{year}{2018}).

\bibitem[{\citenamefont{Starosta et~al.}(2001)\citenamefont{Starosta, Koike,
  Chiara, Fossan, LaFosse, Hecht, Beausang, Caprio, Cooper, Kr\"ucken
  et~al.}}]{Starosta2001PRL}
\bibinfo{author}{\bibfnamefont{K.}~\bibnamefont{Starosta}},
  \bibinfo{author}{\bibfnamefont{T.}~\bibnamefont{Koike}},
  \bibinfo{author}{\bibfnamefont{C.~J.} \bibnamefont{Chiara}},
  \bibinfo{author}{\bibfnamefont{D.~B.} \bibnamefont{Fossan}},
  \bibinfo{author}{\bibfnamefont{D.~R.} \bibnamefont{LaFosse}},
  \bibinfo{author}{\bibfnamefont{A.~A.} \bibnamefont{Hecht}},
  \bibinfo{author}{\bibfnamefont{C.~W.} \bibnamefont{Beausang}},
  \bibinfo{author}{\bibfnamefont{M.~A.} \bibnamefont{Caprio}},
  \bibinfo{author}{\bibfnamefont{J.~R.} \bibnamefont{Cooper}},
  \bibinfo{author}{\bibfnamefont{R.}~\bibnamefont{Kr\"ucken}},
  \bibnamefont{et~al.}, \bibinfo{journal}{Phys. Rev. Lett.}
  \textbf{\bibinfo{volume}{86}}, \bibinfo{pages}{971} (\bibinfo{year}{2001}).

\bibitem[{\citenamefont{Koike et~al.}(2001)\citenamefont{Koike, Starosta,
  Chiara, Fossan, and LaFosse}}]{Koike2001PRC}
\bibinfo{author}{\bibfnamefont{T.}~\bibnamefont{Koike}},
  \bibinfo{author}{\bibfnamefont{K.}~\bibnamefont{Starosta}},
  \bibinfo{author}{\bibfnamefont{C.~J.} \bibnamefont{Chiara}},
  \bibinfo{author}{\bibfnamefont{D.~B.} \bibnamefont{Fossan}},
  \bibnamefont{and} \bibinfo{author}{\bibfnamefont{D.~R.}
  \bibnamefont{LaFosse}}, \bibinfo{journal}{Phys. Rev. C}
  \textbf{\bibinfo{volume}{63}}, \bibinfo{pages}{061304}
  (\bibinfo{year}{2001}).

\bibitem[{\citenamefont{Hecht et~al.}(2001)\citenamefont{Hecht, Beausang,
  Zyromski, Balabanski, Barton, Caprio, Casten, Cooper, Hartley, Krucken
  et~al.}}]{Hecht2001PRC}
\bibinfo{author}{\bibfnamefont{A.~A.} \bibnamefont{Hecht}},
  \bibinfo{author}{\bibfnamefont{C.~W.} \bibnamefont{Beausang}},
  \bibinfo{author}{\bibfnamefont{K.~E.} \bibnamefont{Zyromski}},
  \bibinfo{author}{\bibfnamefont{D.~L.} \bibnamefont{Balabanski}},
  \bibinfo{author}{\bibfnamefont{C.~J.} \bibnamefont{Barton}},
  \bibinfo{author}{\bibfnamefont{M.~A.} \bibnamefont{Caprio}},
  \bibinfo{author}{\bibfnamefont{R.~F.} \bibnamefont{Casten}},
  \bibinfo{author}{\bibfnamefont{J.~R.} \bibnamefont{Cooper}},
  \bibinfo{author}{\bibfnamefont{D.~J.} \bibnamefont{Hartley}},
  \bibinfo{author}{\bibfnamefont{R.}~\bibnamefont{Krucken}},
  \bibnamefont{et~al.}, \bibinfo{journal}{Phys. Rev. C}
  \textbf{\bibinfo{volume}{63}}, \bibinfo{pages}{051302}
  (\bibinfo{year}{2001}).

\bibitem[{\citenamefont{Hartley et~al.}(2001)\citenamefont{Hartley, Riedinger,
  Riley, Balabanski, Kondev, Laird, Pfohl, Archer, Brown, Clark
  et~al.}}]{Hartley2001PRC}
\bibinfo{author}{\bibfnamefont{D.~J.} \bibnamefont{Hartley}},
  \bibinfo{author}{\bibfnamefont{L.~L.} \bibnamefont{Riedinger}},
  \bibinfo{author}{\bibfnamefont{M.~A.} \bibnamefont{Riley}},
  \bibinfo{author}{\bibfnamefont{D.~L.} \bibnamefont{Balabanski}},
  \bibinfo{author}{\bibfnamefont{F.~G.} \bibnamefont{Kondev}},
  \bibinfo{author}{\bibfnamefont{R.~W.} \bibnamefont{Laird}},
  \bibinfo{author}{\bibfnamefont{J.}~\bibnamefont{Pfohl}},
  \bibinfo{author}{\bibfnamefont{D.~E.} \bibnamefont{Archer}},
  \bibinfo{author}{\bibfnamefont{T.~B.} \bibnamefont{Brown}},
  \bibinfo{author}{\bibfnamefont{R.~M.} \bibnamefont{Clark}},
  \bibnamefont{et~al.}, \bibinfo{journal}{Phys. Rev. C}
  \textbf{\bibinfo{volume}{64}}, \bibinfo{pages}{031304}
  (\bibinfo{year}{2001}).

\bibitem[{\citenamefont{Zhu et~al.}(2003)\citenamefont{Zhu, Garg, Nayak,
  Ghugre, Pattabiraman, Fossan, Koike, Starosta, Vaman, Janssens
  et~al.}}]{Zhu2003PRL}
\bibinfo{author}{\bibfnamefont{S.}~\bibnamefont{Zhu}},
  \bibinfo{author}{\bibfnamefont{U.}~\bibnamefont{Garg}},
  \bibinfo{author}{\bibfnamefont{B.~K.} \bibnamefont{Nayak}},
  \bibinfo{author}{\bibfnamefont{S.~S.} \bibnamefont{Ghugre}},
  \bibinfo{author}{\bibfnamefont{N.~S.} \bibnamefont{Pattabiraman}},
  \bibinfo{author}{\bibfnamefont{D.~B.} \bibnamefont{Fossan}},
  \bibinfo{author}{\bibfnamefont{T.}~\bibnamefont{Koike}},
  \bibinfo{author}{\bibfnamefont{K.}~\bibnamefont{Starosta}},
  \bibinfo{author}{\bibfnamefont{C.}~\bibnamefont{Vaman}},
  \bibinfo{author}{\bibfnamefont{R.~V.~F.} \bibnamefont{Janssens}},
  \bibnamefont{et~al.}, \bibinfo{journal}{Phys. Rev. Lett.}
  \textbf{\bibinfo{volume}{91}}, \bibinfo{pages}{132501}
  (\bibinfo{year}{2003}).

\bibitem[{\citenamefont{Koike et~al.}(2003)\citenamefont{Koike, Starosta,
  Chiara, Fossan, and LaFosse}}]{Koike2003PRC}
\bibinfo{author}{\bibfnamefont{T.}~\bibnamefont{Koike}},
  \bibinfo{author}{\bibfnamefont{K.}~\bibnamefont{Starosta}},
  \bibinfo{author}{\bibfnamefont{C.~J.} \bibnamefont{Chiara}},
  \bibinfo{author}{\bibfnamefont{D.~B.} \bibnamefont{Fossan}},
  \bibnamefont{and} \bibinfo{author}{\bibfnamefont{D.~R.}
  \bibnamefont{LaFosse}}, \bibinfo{journal}{Phys. Rev. C}
  \textbf{\bibinfo{volume}{67}}, \bibinfo{pages}{044319}
  (\bibinfo{year}{2003}).

\bibitem[{\citenamefont{Grodner et~al.}(2006)\citenamefont{Grodner, Srebrny,
  Pasternak, Zalewska, Morek, Droste, Mierzejewski, Kowalczyk, Kownacki,
  Kisielinski et~al.}}]{Grodner2006PRL}
\bibinfo{author}{\bibfnamefont{E.}~\bibnamefont{Grodner}},
  \bibinfo{author}{\bibfnamefont{J.}~\bibnamefont{Srebrny}},
  \bibinfo{author}{\bibfnamefont{A.~A.} \bibnamefont{Pasternak}},
  \bibinfo{author}{\bibfnamefont{I.}~\bibnamefont{Zalewska}},
  \bibinfo{author}{\bibfnamefont{T.}~\bibnamefont{Morek}},
  \bibinfo{author}{\bibfnamefont{C.}~\bibnamefont{Droste}},
  \bibinfo{author}{\bibfnamefont{J.}~\bibnamefont{Mierzejewski}},
  \bibinfo{author}{\bibfnamefont{M.}~\bibnamefont{Kowalczyk}},
  \bibinfo{author}{\bibfnamefont{J.}~\bibnamefont{Kownacki}},
  \bibinfo{author}{\bibfnamefont{M.}~\bibnamefont{Kisielinski}},
  \bibnamefont{et~al.}, \bibinfo{journal}{Phys. Rev. Lett.}
  \textbf{\bibinfo{volume}{97}}, \bibinfo{pages}{172501}
  (\bibinfo{year}{2006}).

\bibitem[{\citenamefont{Wang et~al.}(2006)\citenamefont{Wang, Liu, Komatsubara,
  Ma, and Zhang}}]{S.Y.Wang2006PRC}
\bibinfo{author}{\bibfnamefont{S.~Y.} \bibnamefont{Wang}},
  \bibinfo{author}{\bibfnamefont{Y.~Z.} \bibnamefont{Liu}},
  \bibinfo{author}{\bibfnamefont{I.}~\bibnamefont{Komatsubara}},
  \bibinfo{author}{\bibfnamefont{Y.~J.} \bibnamefont{Ma}}, \bibnamefont{and}
  \bibinfo{author}{\bibfnamefont{Y.~H.} \bibnamefont{Zhang}},
  \bibinfo{journal}{Phys. Rev. C} \textbf{\bibinfo{volume}{74}},
  \bibinfo{pages}{017302} (\bibinfo{year}{2006}).

\bibitem[{\citenamefont{Mukhopadhyay et~al.}(2007)\citenamefont{Mukhopadhyay,
  Almehed, Garg, Frauendorf, Li, Rao, Wang, Ghugre, Carpenter, Gros
  et~al.}}]{Mukhopadhyay2007PRL}
\bibinfo{author}{\bibfnamefont{S.}~\bibnamefont{Mukhopadhyay}},
  \bibinfo{author}{\bibfnamefont{D.}~\bibnamefont{Almehed}},
  \bibinfo{author}{\bibfnamefont{U.}~\bibnamefont{Garg}},
  \bibinfo{author}{\bibfnamefont{S.}~\bibnamefont{Frauendorf}},
  \bibinfo{author}{\bibfnamefont{T.}~\bibnamefont{Li}},
  \bibinfo{author}{\bibfnamefont{P.~V.~M.} \bibnamefont{Rao}},
  \bibinfo{author}{\bibfnamefont{X.}~\bibnamefont{Wang}},
  \bibinfo{author}{\bibfnamefont{S.~S.} \bibnamefont{Ghugre}},
  \bibinfo{author}{\bibfnamefont{M.~P.} \bibnamefont{Carpenter}},
  \bibinfo{author}{\bibfnamefont{S.}~\bibnamefont{Gros}}, \bibnamefont{et~al.},
  \bibinfo{journal}{Phys. Rev. Lett.} \textbf{\bibinfo{volume}{99}},
  \bibinfo{pages}{172501} (\bibinfo{year}{2007}).

\bibitem[{\citenamefont{Grodner et~al.}(2011)\citenamefont{Grodner, Sankowska,
  Morek, Rohozinski, Droste, Srebrny, Pasternak, Kisielinski, Kowalczyk,
  Kownacki et~al.}}]{Grodner2011PLB}
\bibinfo{author}{\bibfnamefont{E.}~\bibnamefont{Grodner}},
  \bibinfo{author}{\bibfnamefont{I.}~\bibnamefont{Sankowska}},
  \bibinfo{author}{\bibfnamefont{T.}~\bibnamefont{Morek}},
  \bibinfo{author}{\bibfnamefont{S.~G.} \bibnamefont{Rohozinski}},
  \bibinfo{author}{\bibfnamefont{C.}~\bibnamefont{Droste}},
  \bibinfo{author}{\bibfnamefont{J.}~\bibnamefont{Srebrny}},
  \bibinfo{author}{\bibfnamefont{A.~A.} \bibnamefont{Pasternak}},
  \bibinfo{author}{\bibfnamefont{M.}~\bibnamefont{Kisielinski}},
  \bibinfo{author}{\bibfnamefont{M.}~\bibnamefont{Kowalczyk}},
  \bibinfo{author}{\bibfnamefont{J.}~\bibnamefont{Kownacki}},
  \bibnamefont{et~al.}, \bibinfo{journal}{Phys. Lett. B}
  \textbf{\bibinfo{volume}{703}}, \bibinfo{pages}{46} (\bibinfo{year}{2011}).

\bibitem[{\citenamefont{Ionescu-Bujor et~al.}(2018)\citenamefont{Ionescu-Bujor,
  Aydin, M\ifmmode~\u{a}\else \u{a}\fi{}rginean, Costache, Bucurescu, Florea,
  Glodariu, Ionescu, Iord\ifmmode~\u{a}\else \u{a}\fi{}chescu,
  M\ifmmode~\u{a}\else \u{a}\fi{}rginean et~al.}}]{Bujor2018PRC}
\bibinfo{author}{\bibfnamefont{M.}~\bibnamefont{Ionescu-Bujor}},
  \bibinfo{author}{\bibfnamefont{S.}~\bibnamefont{Aydin}},
  \bibinfo{author}{\bibfnamefont{N.}~\bibnamefont{M\ifmmode~\u{a}\else
  \u{a}\fi{}rginean}},
  \bibinfo{author}{\bibfnamefont{C.}~\bibnamefont{Costache}},
  \bibinfo{author}{\bibfnamefont{D.}~\bibnamefont{Bucurescu}},
  \bibinfo{author}{\bibfnamefont{N.}~\bibnamefont{Florea}},
  \bibinfo{author}{\bibfnamefont{T.}~\bibnamefont{Glodariu}},
  \bibinfo{author}{\bibfnamefont{A.}~\bibnamefont{Ionescu}},
  \bibinfo{author}{\bibfnamefont{A.}~\bibnamefont{Iord\ifmmode~\u{a}\else
  \u{a}\fi{}chescu}},
  \bibinfo{author}{\bibfnamefont{R.}~\bibnamefont{M\ifmmode~\u{a}\else
  \u{a}\fi{}rginean}}, \bibnamefont{et~al.}, \bibinfo{journal}{Phys. Rev. C}
  \textbf{\bibinfo{volume}{98}}, \bibinfo{pages}{054305}
  (\bibinfo{year}{2018}).

\bibitem[{\citenamefont{Balabanski et~al.}(2004)\citenamefont{Balabanski,
  Danchev, Hartley, Riedinger, Zeidan, Zhang, Barton, Beausang, Caprio, Casten
  et~al.}}]{Balabanski2004PRC}
\bibinfo{author}{\bibfnamefont{D.~L.} \bibnamefont{Balabanski}},
  \bibinfo{author}{\bibfnamefont{M.}~\bibnamefont{Danchev}},
  \bibinfo{author}{\bibfnamefont{D.~J.} \bibnamefont{Hartley}},
  \bibinfo{author}{\bibfnamefont{L.~L.} \bibnamefont{Riedinger}},
  \bibinfo{author}{\bibfnamefont{O.}~\bibnamefont{Zeidan}},
  \bibinfo{author}{\bibfnamefont{J.-y.} \bibnamefont{Zhang}},
  \bibinfo{author}{\bibfnamefont{C.~J.} \bibnamefont{Barton}},
  \bibinfo{author}{\bibfnamefont{C.~W.} \bibnamefont{Beausang}},
  \bibinfo{author}{\bibfnamefont{M.~A.} \bibnamefont{Caprio}},
  \bibinfo{author}{\bibfnamefont{R.~F.} \bibnamefont{Casten}},
  \bibnamefont{et~al.}, \bibinfo{journal}{Phys. Rev. C}
  \textbf{\bibinfo{volume}{70}}, \bibinfo{pages}{044305}
  (\bibinfo{year}{2004}).

\bibitem[{\citenamefont{Lawrie et~al.}(2008)\citenamefont{Lawrie, Vymers,
  Lawrie, Vieu, Bark, Lindsay, Mabala, Maliage, Masiteng, Mullins
  et~al.}}]{Lawrie2008PRC}
\bibinfo{author}{\bibfnamefont{E.~A.} \bibnamefont{Lawrie}},
  \bibinfo{author}{\bibfnamefont{P.~A.} \bibnamefont{Vymers}},
  \bibinfo{author}{\bibfnamefont{J.~J.} \bibnamefont{Lawrie}},
  \bibinfo{author}{\bibfnamefont{C.}~\bibnamefont{Vieu}},
  \bibinfo{author}{\bibfnamefont{R.~A.} \bibnamefont{Bark}},
  \bibinfo{author}{\bibfnamefont{R.}~\bibnamefont{Lindsay}},
  \bibinfo{author}{\bibfnamefont{G.~K.} \bibnamefont{Mabala}},
  \bibinfo{author}{\bibfnamefont{S.~M.} \bibnamefont{Maliage}},
  \bibinfo{author}{\bibfnamefont{P.~L.} \bibnamefont{Masiteng}},
  \bibinfo{author}{\bibfnamefont{S.~M.} \bibnamefont{Mullins}},
  \bibnamefont{et~al.}, \bibinfo{journal}{Phys. Rev. C}
  \textbf{\bibinfo{volume}{78}}, \bibinfo{pages}{021305}
  (\bibinfo{year}{2008}).

\bibitem[{\citenamefont{Meng and Zhang}(2010)}]{J.Meng2010JPG}
\bibinfo{author}{\bibfnamefont{J.}~\bibnamefont{Meng}} \bibnamefont{and}
  \bibinfo{author}{\bibfnamefont{S.~Q.} \bibnamefont{Zhang}},
  \bibinfo{journal}{J. Phys. G: Nucl. Part. Phys.}
  \textbf{\bibinfo{volume}{37}}, \bibinfo{pages}{064025}
  (\bibinfo{year}{2010}).

\bibitem[{\citenamefont{Meng et~al.}(2014)\citenamefont{Meng, Chen, and
  Zhang}}]{J.Meng2014IJMPE}
\bibinfo{author}{\bibfnamefont{J.}~\bibnamefont{Meng}},
  \bibinfo{author}{\bibfnamefont{Q.~B.} \bibnamefont{Chen}}, \bibnamefont{and}
  \bibinfo{author}{\bibfnamefont{S.~Q.} \bibnamefont{Zhang}},
  \bibinfo{journal}{Int. J. Mod. Phys. E} \textbf{\bibinfo{volume}{23}},
  \bibinfo{pages}{1430016} (\bibinfo{year}{2014}).

\bibitem[{\citenamefont{Bark et~al.}(2014)\citenamefont{Bark, Lieder, Lieder,
  Lawrie, Lawrie, Bvumbi, Kheswa, Ntshangase, Madiba, Masiteng
  et~al.}}]{Bark2014IJMPE}
\bibinfo{author}{\bibfnamefont{R.~A.} \bibnamefont{Bark}},
  \bibinfo{author}{\bibfnamefont{E.~O.} \bibnamefont{Lieder}},
  \bibinfo{author}{\bibfnamefont{R.~M.} \bibnamefont{Lieder}},
  \bibinfo{author}{\bibfnamefont{E.~A.} \bibnamefont{Lawrie}},
  \bibinfo{author}{\bibfnamefont{J.~J.} \bibnamefont{Lawrie}},
  \bibinfo{author}{\bibfnamefont{S.~P.} \bibnamefont{Bvumbi}},
  \bibinfo{author}{\bibfnamefont{N.~Y.} \bibnamefont{Kheswa}},
  \bibinfo{author}{\bibfnamefont{S.~S.} \bibnamefont{Ntshangase}},
  \bibinfo{author}{\bibfnamefont{T.~E.} \bibnamefont{Madiba}},
  \bibinfo{author}{\bibfnamefont{P.~L.} \bibnamefont{Masiteng}},
  \bibnamefont{et~al.}, \bibinfo{journal}{Int. J. Mod. Phys. E}
  \textbf{\bibinfo{volume}{23}}, \bibinfo{pages}{1461001}
  (\bibinfo{year}{2014}).

\bibitem[{\citenamefont{Meng and Zhao}(2016)}]{J.Meng2016PS}
\bibinfo{author}{\bibfnamefont{J.}~\bibnamefont{Meng}} \bibnamefont{and}
  \bibinfo{author}{\bibfnamefont{P.~W.} \bibnamefont{Zhao}},
  \bibinfo{journal}{Phys. Scr.} \textbf{\bibinfo{volume}{91}},
  \bibinfo{pages}{053008} (\bibinfo{year}{2016}).

\bibitem[{\citenamefont{Raduta}(2016)}]{Raduta2016PPNP}
\bibinfo{author}{\bibfnamefont{A.~A.} \bibnamefont{Raduta}},
  \bibinfo{journal}{Prog. Part. Nucl. Phys.} \textbf{\bibinfo{volume}{90}},
  \bibinfo{pages}{241} (\bibinfo{year}{2016}).

\bibitem[{\citenamefont{Frauendorf}(2018)}]{Frauendorf2018PS}
\bibinfo{author}{\bibfnamefont{S.}~\bibnamefont{Frauendorf}},
  \bibinfo{journal}{Phys. Scr.} \textbf{\bibinfo{volume}{93}},
  \bibinfo{pages}{043003} (\bibinfo{year}{2018}).

\bibitem[{\citenamefont{Xiong and Wang}(2019)}]{B.W.Xiong2019ADNDT}
\bibinfo{author}{\bibfnamefont{B.~W.} \bibnamefont{Xiong}} \bibnamefont{and}
  \bibinfo{author}{\bibfnamefont{Y.~Y.} \bibnamefont{Wang}},
  \bibinfo{journal}{Atom. Data Nucl. Data Tables}
  \textbf{\bibinfo{volume}{125}}, \bibinfo{pages}{193} (\bibinfo{year}{2019}).

\bibitem[{\citenamefont{Meng et~al.}(2006)\citenamefont{Meng, Peng, Zhang, and
  Zhou}}]{J.Meng2006PRC}
\bibinfo{author}{\bibfnamefont{J.}~\bibnamefont{Meng}},
  \bibinfo{author}{\bibfnamefont{J.}~\bibnamefont{Peng}},
  \bibinfo{author}{\bibfnamefont{S.~Q.} \bibnamefont{Zhang}}, \bibnamefont{and}
  \bibinfo{author}{\bibfnamefont{S.-G.} \bibnamefont{Zhou}},
  \bibinfo{journal}{Phys. Rev. C} \textbf{\bibinfo{volume}{73}},
  \bibinfo{pages}{037303} (\bibinfo{year}{2006}).

\bibitem[{\citenamefont{Peng et~al.}(2008)\citenamefont{Peng, Sagawa, Zhang,
  Yao, Zhang, and Meng}}]{J.Peng2008PRC}
\bibinfo{author}{\bibfnamefont{J.}~\bibnamefont{Peng}},
  \bibinfo{author}{\bibfnamefont{H.}~\bibnamefont{Sagawa}},
  \bibinfo{author}{\bibfnamefont{S.~Q.} \bibnamefont{Zhang}},
  \bibinfo{author}{\bibfnamefont{J.~M.} \bibnamefont{Yao}},
  \bibinfo{author}{\bibfnamefont{Y.}~\bibnamefont{Zhang}}, \bibnamefont{and}
  \bibinfo{author}{\bibfnamefont{J.}~\bibnamefont{Meng}},
  \bibinfo{journal}{Phys. Rev. C} \textbf{\bibinfo{volume}{77}},
  \bibinfo{pages}{024309} (\bibinfo{year}{2008}).

\bibitem[{\citenamefont{Yao et~al.}(2009)\citenamefont{Yao, Qi, Zhang, Peng,
  Wang, and Meng}}]{J.M.Yao2009PRC}
\bibinfo{author}{\bibfnamefont{J.~M.} \bibnamefont{Yao}},
  \bibinfo{author}{\bibfnamefont{B.}~\bibnamefont{Qi}},
  \bibinfo{author}{\bibfnamefont{S.~Q.} \bibnamefont{Zhang}},
  \bibinfo{author}{\bibfnamefont{J.}~\bibnamefont{Peng}},
  \bibinfo{author}{\bibfnamefont{S.~Y.} \bibnamefont{Wang}}, \bibnamefont{and}
  \bibinfo{author}{\bibfnamefont{J.}~\bibnamefont{Meng}},
  \bibinfo{journal}{Phys. Rev. C} \textbf{\bibinfo{volume}{79}},
  \bibinfo{pages}{067302} (\bibinfo{year}{2009}).

\bibitem[{\citenamefont{Li et~al.}(2011)\citenamefont{Li, Zhang, and
  Meng}}]{J.Li2011PRC}
\bibinfo{author}{\bibfnamefont{J.}~\bibnamefont{Li}},
  \bibinfo{author}{\bibfnamefont{S.~Q.} \bibnamefont{Zhang}}, \bibnamefont{and}
  \bibinfo{author}{\bibfnamefont{J.}~\bibnamefont{Meng}},
  \bibinfo{journal}{Phys. Rev. C} \textbf{\bibinfo{volume}{83}},
  \bibinfo{pages}{037301} (\bibinfo{year}{2011}).

\bibitem[{\citenamefont{Li}(2018)}]{J.Li2018PRC}
\bibinfo{author}{\bibfnamefont{J.}~\bibnamefont{Li}}, \bibinfo{journal}{Phys.
  Rev. C} \textbf{\bibinfo{volume}{97}}, \bibinfo{pages}{034306}
  (\bibinfo{year}{2018}).

\bibitem[{\citenamefont{Qi et~al.}(2018)\citenamefont{Qi, Jia, Liu, and
  Wang}}]{B.Qi2018PRC}
\bibinfo{author}{\bibfnamefont{B.}~\bibnamefont{Qi}},
  \bibinfo{author}{\bibfnamefont{H.}~\bibnamefont{Jia}},
  \bibinfo{author}{\bibfnamefont{C.}~\bibnamefont{Liu}}, \bibnamefont{and}
  \bibinfo{author}{\bibfnamefont{S.~Y.} \bibnamefont{Wang}},
  \bibinfo{journal}{Phys. Rev. C} \textbf{\bibinfo{volume}{98}},
  \bibinfo{pages}{014305} (\bibinfo{year}{2018}).

\bibitem[{\citenamefont{Peng and Chen}(2018)}]{J.Peng2018PRC}
\bibinfo{author}{\bibfnamefont{J.}~\bibnamefont{Peng}} \bibnamefont{and}
  \bibinfo{author}{\bibfnamefont{Q.~B.} \bibnamefont{Chen}},
  \bibinfo{journal}{Phys. Rev. C} \textbf{\bibinfo{volume}{98}},
  \bibinfo{pages}{024320} (\bibinfo{year}{2018}).

\bibitem[{\citenamefont{Ayangeakaa et~al.}(2013)\citenamefont{Ayangeakaa, Garg,
  Anthony, Frauendorf, Matta, Nayak, Patel, Chen, Zhang, Zhao
  et~al.}}]{Ayangeakaa2013PRL}
\bibinfo{author}{\bibfnamefont{A.~D.} \bibnamefont{Ayangeakaa}},
  \bibinfo{author}{\bibfnamefont{U.}~\bibnamefont{Garg}},
  \bibinfo{author}{\bibfnamefont{M.~D.} \bibnamefont{Anthony}},
  \bibinfo{author}{\bibfnamefont{S.}~\bibnamefont{Frauendorf}},
  \bibinfo{author}{\bibfnamefont{J.~T.} \bibnamefont{Matta}},
  \bibinfo{author}{\bibfnamefont{B.~K.} \bibnamefont{Nayak}},
  \bibinfo{author}{\bibfnamefont{D.}~\bibnamefont{Patel}},
  \bibinfo{author}{\bibfnamefont{Q.~B.} \bibnamefont{Chen}},
  \bibinfo{author}{\bibfnamefont{S.~Q.} \bibnamefont{Zhang}},
  \bibinfo{author}{\bibfnamefont{P.~W.} \bibnamefont{Zhao}},
  \bibnamefont{et~al.}, \bibinfo{journal}{Phys. Rev. Lett.}
  \textbf{\bibinfo{volume}{110}}, \bibinfo{pages}{172504}
  (\bibinfo{year}{2013}).

\bibitem[{\citenamefont{Kuti et~al.}(2014)\citenamefont{Kuti, Chen, Tim\'ar,
  Sohler, Zhang, Zhang, Zhao, Meng, Starosta, Koike et~al.}}]{Kuti2014PRL}
\bibinfo{author}{\bibfnamefont{I.}~\bibnamefont{Kuti}},
  \bibinfo{author}{\bibfnamefont{Q.~B.} \bibnamefont{Chen}},
  \bibinfo{author}{\bibfnamefont{J.}~\bibnamefont{Tim\'ar}},
  \bibinfo{author}{\bibfnamefont{D.}~\bibnamefont{Sohler}},
  \bibinfo{author}{\bibfnamefont{S.~Q.} \bibnamefont{Zhang}},
  \bibinfo{author}{\bibfnamefont{Z.~H.} \bibnamefont{Zhang}},
  \bibinfo{author}{\bibfnamefont{P.~W.} \bibnamefont{Zhao}},
  \bibinfo{author}{\bibfnamefont{J.}~\bibnamefont{Meng}},
  \bibinfo{author}{\bibfnamefont{K.}~\bibnamefont{Starosta}},
  \bibinfo{author}{\bibfnamefont{T.}~\bibnamefont{Koike}},
  \bibnamefont{et~al.}, \bibinfo{journal}{Phys. Rev. Lett.}
  \textbf{\bibinfo{volume}{113}}, \bibinfo{pages}{032501}
  (\bibinfo{year}{2014}).

\bibitem[{\citenamefont{Liu et~al.}(2016)\citenamefont{Liu, Wang, Bark, Zhang,
  Meng, Qi, Jones, Wyngaardt, Zhao, Xu et~al.}}]{C.Liu2016PRL}
\bibinfo{author}{\bibfnamefont{C.}~\bibnamefont{Liu}},
  \bibinfo{author}{\bibfnamefont{S.~Y.} \bibnamefont{Wang}},
  \bibinfo{author}{\bibfnamefont{R.~A.} \bibnamefont{Bark}},
  \bibinfo{author}{\bibfnamefont{S.~Q.} \bibnamefont{Zhang}},
  \bibinfo{author}{\bibfnamefont{J.}~\bibnamefont{Meng}},
  \bibinfo{author}{\bibfnamefont{B.}~\bibnamefont{Qi}},
  \bibinfo{author}{\bibfnamefont{P.}~\bibnamefont{Jones}},
  \bibinfo{author}{\bibfnamefont{S.~M.} \bibnamefont{Wyngaardt}},
  \bibinfo{author}{\bibfnamefont{J.}~\bibnamefont{Zhao}},
  \bibinfo{author}{\bibfnamefont{C.}~\bibnamefont{Xu}}, \bibnamefont{et~al.},
  \bibinfo{journal}{Phys. Rev. Lett.} \textbf{\bibinfo{volume}{116}},
  \bibinfo{pages}{112501} (\bibinfo{year}{2016}).

\bibitem[{\citenamefont{Petrache et~al.}(2018)\citenamefont{Petrache, Lv,
  Astier, Dupont, Wang, Zhang, Zhao, Ren, Meng, Greenlees
  et~al.}}]{Petrache2018PRC}
\bibinfo{author}{\bibfnamefont{C.~M.} \bibnamefont{Petrache}},
  \bibinfo{author}{\bibfnamefont{B.~F.} \bibnamefont{Lv}},
  \bibinfo{author}{\bibfnamefont{A.}~\bibnamefont{Astier}},
  \bibinfo{author}{\bibfnamefont{E.}~\bibnamefont{Dupont}},
  \bibinfo{author}{\bibfnamefont{Y.~K.} \bibnamefont{Wang}},
  \bibinfo{author}{\bibfnamefont{S.~Q.} \bibnamefont{Zhang}},
  \bibinfo{author}{\bibfnamefont{P.~W.} \bibnamefont{Zhao}},
  \bibinfo{author}{\bibfnamefont{Z.~X.} \bibnamefont{Ren}},
  \bibinfo{author}{\bibfnamefont{J.}~\bibnamefont{Meng}},
  \bibinfo{author}{\bibfnamefont{P.~T.} \bibnamefont{Greenlees}},
  \bibnamefont{et~al.}, \bibinfo{journal}{Phys. Rev. C}
  \textbf{\bibinfo{volume}{97}}, \bibinfo{pages}{041304(R)}
  (\bibinfo{year}{2018}).

\bibitem[{\citenamefont{Chen et~al.}(2018{\natexlab{a}})\citenamefont{Chen, Lv,
  Petrache, and Meng}}]{Q.B.Chen2018PLB}
\bibinfo{author}{\bibfnamefont{Q.~B.} \bibnamefont{Chen}},
  \bibinfo{author}{\bibfnamefont{B.~F.} \bibnamefont{Lv}},
  \bibinfo{author}{\bibfnamefont{C.~M.} \bibnamefont{Petrache}},
  \bibnamefont{and} \bibinfo{author}{\bibfnamefont{J.}~\bibnamefont{Meng}},
  \bibinfo{journal}{Phys. Lett. B} \textbf{\bibinfo{volume}{782}},
  \bibinfo{pages}{744} (\bibinfo{year}{2018}{\natexlab{a}}).

\bibitem[{\citenamefont{Roy et~al.}(2018)\citenamefont{Roy, Mukherjee, Asgar,
  Bhattacharyya, Bhattacharya, Bhattacharya, Bhattacharya, Ghosh, Banerjee,
  Kundu et~al.}}]{Roy2018PLB}
\bibinfo{author}{\bibfnamefont{T.}~\bibnamefont{Roy}},
  \bibinfo{author}{\bibfnamefont{G.}~\bibnamefont{Mukherjee}},
  \bibinfo{author}{\bibfnamefont{M.}~\bibnamefont{Asgar}},
  \bibinfo{author}{\bibfnamefont{S.}~\bibnamefont{Bhattacharyya}},
  \bibinfo{author}{\bibfnamefont{S.}~\bibnamefont{Bhattacharya}},
  \bibinfo{author}{\bibfnamefont{C.}~\bibnamefont{Bhattacharya}},
  \bibinfo{author}{\bibfnamefont{S.}~\bibnamefont{Bhattacharya}},
  \bibinfo{author}{\bibfnamefont{T.}~\bibnamefont{Ghosh}},
  \bibinfo{author}{\bibfnamefont{K.}~\bibnamefont{Banerjee}},
  \bibinfo{author}{\bibfnamefont{S.}~\bibnamefont{Kundu}},
  \bibnamefont{et~al.}, \bibinfo{journal}{Phys. Lett. B}
  \textbf{\bibinfo{volume}{782}}, \bibinfo{pages}{768 } (\bibinfo{year}{2018}).

\bibitem[{\citenamefont{Droste et~al.}(2009)\citenamefont{Droste, Rohozinski,
  Starosta, Prochniak, and Grodner}}]{Droste2009EPJA}
\bibinfo{author}{\bibfnamefont{C.}~\bibnamefont{Droste}},
  \bibinfo{author}{\bibfnamefont{S.~G.} \bibnamefont{Rohozinski}},
  \bibinfo{author}{\bibfnamefont{K.}~\bibnamefont{Starosta}},
  \bibinfo{author}{\bibfnamefont{L.}~\bibnamefont{Prochniak}},
  \bibnamefont{and} \bibinfo{author}{\bibfnamefont{E.}~\bibnamefont{Grodner}},
  \bibinfo{journal}{Eur. Phys. J. A} \textbf{\bibinfo{volume}{42}},
  \bibinfo{pages}{79} (\bibinfo{year}{2009}).

\bibitem[{\citenamefont{Chen et~al.}(2010)\citenamefont{Chen, Yao, Zhang, and
  Qi}}]{Q.B.Chen2010PRC}
\bibinfo{author}{\bibfnamefont{Q.~B.} \bibnamefont{Chen}},
  \bibinfo{author}{\bibfnamefont{J.~M.} \bibnamefont{Yao}},
  \bibinfo{author}{\bibfnamefont{S.~Q.} \bibnamefont{Zhang}}, \bibnamefont{and}
  \bibinfo{author}{\bibfnamefont{B.}~\bibnamefont{Qi}}, \bibinfo{journal}{Phys.
  Rev. C} \textbf{\bibinfo{volume}{82}}, \bibinfo{pages}{067302}
  (\bibinfo{year}{2010}).

\bibitem[{\citenamefont{Hamamoto}(2013)}]{Hamamoto2013PRC}
\bibinfo{author}{\bibfnamefont{I.}~\bibnamefont{Hamamoto}},
  \bibinfo{journal}{Phys. Rev. C} \textbf{\bibinfo{volume}{88}},
  \bibinfo{pages}{024327} (\bibinfo{year}{2013}).

\bibitem[{\citenamefont{Zhang and Chen}(2016)}]{H.Zhang2016CPC}
\bibinfo{author}{\bibfnamefont{H.}~\bibnamefont{Zhang}} \bibnamefont{and}
  \bibinfo{author}{\bibfnamefont{Q.~B.} \bibnamefont{Chen}},
  \bibinfo{journal}{Chin. Phys. C} \textbf{\bibinfo{volume}{40}},
  \bibinfo{pages}{024101} (\bibinfo{year}{2016}).

\bibitem[{\citenamefont{Zhao et~al.}(2011)\citenamefont{Zhao, Zhang, Peng,
  Liang, Ring, and Meng}}]{P.W.Zhao2011PLB}
\bibinfo{author}{\bibfnamefont{P.~W.} \bibnamefont{Zhao}},
  \bibinfo{author}{\bibfnamefont{S.~Q.} \bibnamefont{Zhang}},
  \bibinfo{author}{\bibfnamefont{J.}~\bibnamefont{Peng}},
  \bibinfo{author}{\bibfnamefont{H.~Z.} \bibnamefont{Liang}},
  \bibinfo{author}{\bibfnamefont{P.}~\bibnamefont{Ring}}, \bibnamefont{and}
  \bibinfo{author}{\bibfnamefont{J.}~\bibnamefont{Meng}},
  \bibinfo{journal}{Phys. Lett. B} \textbf{\bibinfo{volume}{699}},
  \bibinfo{pages}{181} (\bibinfo{year}{2011}).

\bibitem[{\citenamefont{Torres et~al.}(2008)\citenamefont{Torres, Cristancho,
  Andersson, Johansson, Rudolph, Fahlander, Ekman, du~Rietz, Andreoiu,
  Carpenter et~al.}}]{Torres2008PRC}
\bibinfo{author}{\bibfnamefont{D.~A.} \bibnamefont{Torres}},
  \bibinfo{author}{\bibfnamefont{F.}~\bibnamefont{Cristancho}},
  \bibinfo{author}{\bibfnamefont{L.-L.} \bibnamefont{Andersson}},
  \bibinfo{author}{\bibfnamefont{E.~K.} \bibnamefont{Johansson}},
  \bibinfo{author}{\bibfnamefont{D.}~\bibnamefont{Rudolph}},
  \bibinfo{author}{\bibfnamefont{C.}~\bibnamefont{Fahlander}},
  \bibinfo{author}{\bibfnamefont{J.}~\bibnamefont{Ekman}},
  \bibinfo{author}{\bibfnamefont{R.}~\bibnamefont{du~Rietz}},
  \bibinfo{author}{\bibfnamefont{C.}~\bibnamefont{Andreoiu}},
  \bibinfo{author}{\bibfnamefont{M.~P.} \bibnamefont{Carpenter}},
  \bibnamefont{et~al.}, \bibinfo{journal}{Phys. Rev. C}
  \textbf{\bibinfo{volume}{78}}, \bibinfo{pages}{054318}
  (\bibinfo{year}{2008}).

\bibitem[{\citenamefont{Meng}(2016)}]{J.Meng2016book}
\bibinfo{editor}{\bibfnamefont{J.}~\bibnamefont{Meng}}, ed.,
  \emph{\bibinfo{title}{Relativistic density functional for nuclear
  structure}}, vol.~\bibinfo{volume}{10} of
  \emph{\bibinfo{series}{International Review of Nuclear Physics}}
  (\bibinfo{publisher}{World Scientific}, \bibinfo{address}{Singapore},
  \bibinfo{year}{2016}).

\bibitem[{\citenamefont{Petrache et~al.}(2016)\citenamefont{Petrache, Chen,
  Guo, Ayangeakaa, Garg, Matta, Nayak, Patel, Meng, Carpenter
  et~al.}}]{Petrache2016PRC}
\bibinfo{author}{\bibfnamefont{C.~M.} \bibnamefont{Petrache}},
  \bibinfo{author}{\bibfnamefont{Q.~B.} \bibnamefont{Chen}},
  \bibinfo{author}{\bibfnamefont{S.}~\bibnamefont{Guo}},
  \bibinfo{author}{\bibfnamefont{A.~D.} \bibnamefont{Ayangeakaa}},
  \bibinfo{author}{\bibfnamefont{U.}~\bibnamefont{Garg}},
  \bibinfo{author}{\bibfnamefont{J.~T.} \bibnamefont{Matta}},
  \bibinfo{author}{\bibfnamefont{B.~K.} \bibnamefont{Nayak}},
  \bibinfo{author}{\bibfnamefont{D.}~\bibnamefont{Patel}},
  \bibinfo{author}{\bibfnamefont{J.}~\bibnamefont{Meng}},
  \bibinfo{author}{\bibfnamefont{M.~P.} \bibnamefont{Carpenter}},
  \bibnamefont{et~al.}, \bibinfo{journal}{Phys. Rev. C}
  \textbf{\bibinfo{volume}{94}}, \bibinfo{pages}{064309}
  (\bibinfo{year}{2016}).

\bibitem[{\citenamefont{Peng et~al.}(2003)\citenamefont{Peng, Meng, and
  Zhang}}]{J.Peng2003PRC}
\bibinfo{author}{\bibfnamefont{J.}~\bibnamefont{Peng}},
  \bibinfo{author}{\bibfnamefont{J.}~\bibnamefont{Meng}}, \bibnamefont{and}
  \bibinfo{author}{\bibfnamefont{S.~Q.} \bibnamefont{Zhang}},
  \bibinfo{journal}{Phys. Rev. C} \textbf{\bibinfo{volume}{68}},
  \bibinfo{pages}{044324} (\bibinfo{year}{2003}).

\bibitem[{\citenamefont{Koike et~al.}(2004)\citenamefont{Koike, Starosta, and
  Hamamoto}}]{Koike2004PRL}
\bibinfo{author}{\bibfnamefont{T.}~\bibnamefont{Koike}},
  \bibinfo{author}{\bibfnamefont{K.}~\bibnamefont{Starosta}}, \bibnamefont{and}
  \bibinfo{author}{\bibfnamefont{I.}~\bibnamefont{Hamamoto}},
  \bibinfo{journal}{Phys. Rev. Lett.} \textbf{\bibinfo{volume}{93}},
  \bibinfo{pages}{172502} (\bibinfo{year}{2004}).

\bibitem[{\citenamefont{Qi et~al.}(2009{\natexlab{a}})\citenamefont{Qi, Zhang,
  Wang, Yao, and Meng}}]{B.Qi2009PRC}
\bibinfo{author}{\bibfnamefont{B.}~\bibnamefont{Qi}},
  \bibinfo{author}{\bibfnamefont{S.~Q.} \bibnamefont{Zhang}},
  \bibinfo{author}{\bibfnamefont{S.~Y.} \bibnamefont{Wang}},
  \bibinfo{author}{\bibfnamefont{J.~M.} \bibnamefont{Yao}}, \bibnamefont{and}
  \bibinfo{author}{\bibfnamefont{J.}~\bibnamefont{Meng}},
  \bibinfo{journal}{Phys. Rev. C} \textbf{\bibinfo{volume}{79}},
  \bibinfo{pages}{041302(R)} (\bibinfo{year}{2009}{\natexlab{a}}).

\bibitem[{\citenamefont{Chen et~al.}(2018{\natexlab{b}})\citenamefont{Chen,
  Starosta, and Koike}}]{Q.B.Chen2018PRC}
\bibinfo{author}{\bibfnamefont{Q.~B.} \bibnamefont{Chen}},
  \bibinfo{author}{\bibfnamefont{K.}~\bibnamefont{Starosta}}, \bibnamefont{and}
  \bibinfo{author}{\bibfnamefont{T.}~\bibnamefont{Koike}},
  \bibinfo{journal}{Phys. Rev. C} \textbf{\bibinfo{volume}{97}},
  \bibinfo{pages}{041303(R)} (\bibinfo{year}{2018}{\natexlab{b}}).

\bibitem[{\citenamefont{Chen and Meng}(2018)}]{Q.B.Chen2018PRC_v1}
\bibinfo{author}{\bibfnamefont{Q.~B.} \bibnamefont{Chen}} \bibnamefont{and}
  \bibinfo{author}{\bibfnamefont{J.}~\bibnamefont{Meng}},
  \bibinfo{journal}{Phys. Rev. C} \textbf{\bibinfo{volume}{98}},
  \bibinfo{pages}{031303} (\bibinfo{year}{2018}).

\bibitem[{\citenamefont{Zhang et~al.}(2007)\citenamefont{Zhang, Qi, Wang, and
  Meng}}]{S.Q.Zhang2007PRC}
\bibinfo{author}{\bibfnamefont{S.~Q.} \bibnamefont{Zhang}},
  \bibinfo{author}{\bibfnamefont{B.}~\bibnamefont{Qi}},
  \bibinfo{author}{\bibfnamefont{S.~Y.} \bibnamefont{Wang}}, \bibnamefont{and}
  \bibinfo{author}{\bibfnamefont{J.}~\bibnamefont{Meng}},
  \bibinfo{journal}{Phys. Rev. C} \textbf{\bibinfo{volume}{75}},
  \bibinfo{pages}{044307} (\bibinfo{year}{2007}).

\bibitem[{\citenamefont{Wang et~al.}(2007)\citenamefont{Wang, Zhang, Qi, and
  Meng}}]{S.Y.Wang2007PRC}
\bibinfo{author}{\bibfnamefont{S.~Y.} \bibnamefont{Wang}},
  \bibinfo{author}{\bibfnamefont{S.~Q.} \bibnamefont{Zhang}},
  \bibinfo{author}{\bibfnamefont{B.}~\bibnamefont{Qi}}, \bibnamefont{and}
  \bibinfo{author}{\bibfnamefont{J.}~\bibnamefont{Meng}},
  \bibinfo{journal}{Phys. Rev. C} \textbf{\bibinfo{volume}{75}},
  \bibinfo{pages}{024309} (\bibinfo{year}{2007}).

\bibitem[{\citenamefont{Wang et~al.}(2008)\citenamefont{Wang, Zhang, Qi, Peng,
  Yao, and Meng}}]{S.Y.Wang2008PRC}
\bibinfo{author}{\bibfnamefont{S.~Y.} \bibnamefont{Wang}},
  \bibinfo{author}{\bibfnamefont{S.~Q.} \bibnamefont{Zhang}},
  \bibinfo{author}{\bibfnamefont{B.}~\bibnamefont{Qi}},
  \bibinfo{author}{\bibfnamefont{J.}~\bibnamefont{Peng}},
  \bibinfo{author}{\bibfnamefont{J.~M.} \bibnamefont{Yao}}, \bibnamefont{and}
  \bibinfo{author}{\bibfnamefont{J.}~\bibnamefont{Meng}},
  \bibinfo{journal}{Phys. Rev. C} \textbf{\bibinfo{volume}{77}},
  \bibinfo{pages}{034314} (\bibinfo{year}{2008}).

\bibitem[{\citenamefont{Lawrie and Shirinda}(2010)}]{Lawrie2010PLB}
\bibinfo{author}{\bibfnamefont{E.~A.} \bibnamefont{Lawrie}} \bibnamefont{and}
  \bibinfo{author}{\bibfnamefont{O.}~\bibnamefont{Shirinda}},
  \bibinfo{journal}{Phys. Lett. B} \textbf{\bibinfo{volume}{689}},
  \bibinfo{pages}{66} (\bibinfo{year}{2010}).

\bibitem[{\citenamefont{Shirinda and Lawrie}(2012)}]{Shirinda2012EPJA}
\bibinfo{author}{\bibfnamefont{O.}~\bibnamefont{Shirinda}} \bibnamefont{and}
  \bibinfo{author}{\bibfnamefont{E.~A.} \bibnamefont{Lawrie}},
  \bibinfo{journal}{Eur. Phys. J. A} \textbf{\bibinfo{volume}{48}},
  \bibinfo{pages}{118} (\bibinfo{year}{2012}).

\bibitem[{\citenamefont{Qi et~al.}(2009{\natexlab{b}})\citenamefont{Qi, Zhang,
  Meng, Wang, and Frauendorf}}]{B.Qi2009PLB}
\bibinfo{author}{\bibfnamefont{B.}~\bibnamefont{Qi}},
  \bibinfo{author}{\bibfnamefont{S.~Q.} \bibnamefont{Zhang}},
  \bibinfo{author}{\bibfnamefont{J.}~\bibnamefont{Meng}},
  \bibinfo{author}{\bibfnamefont{S.~Y.} \bibnamefont{Wang}}, \bibnamefont{and}
  \bibinfo{author}{\bibfnamefont{S.}~\bibnamefont{Frauendorf}},
  \bibinfo{journal}{Phys. Lett. B} \textbf{\bibinfo{volume}{675}},
  \bibinfo{pages}{175} (\bibinfo{year}{2009}{\natexlab{b}}).

\bibitem[{\citenamefont{Qi et~al.}(2011)\citenamefont{Qi, Zhang, Wang, Meng,
  and Koike}}]{B.Qi2011PRC}
\bibinfo{author}{\bibfnamefont{B.}~\bibnamefont{Qi}},
  \bibinfo{author}{\bibfnamefont{S.~Q.} \bibnamefont{Zhang}},
  \bibinfo{author}{\bibfnamefont{S.~Y.} \bibnamefont{Wang}},
  \bibinfo{author}{\bibfnamefont{J.}~\bibnamefont{Meng}}, \bibnamefont{and}
  \bibinfo{author}{\bibfnamefont{T.}~\bibnamefont{Koike}},
  \bibinfo{journal}{Phys. Rev. C} \textbf{\bibinfo{volume}{83}},
  \bibinfo{pages}{034303} (\bibinfo{year}{2011}).

\bibitem[{\citenamefont{Qi et~al.}(2013)\citenamefont{Qi, Jia, Zhang, Liu, and
  Wang}}]{B.Qi2013PRC}
\bibinfo{author}{\bibfnamefont{B.}~\bibnamefont{Qi}},
  \bibinfo{author}{\bibfnamefont{H.}~\bibnamefont{Jia}},
  \bibinfo{author}{\bibfnamefont{N.~B.} \bibnamefont{Zhang}},
  \bibinfo{author}{\bibfnamefont{C.}~\bibnamefont{Liu}}, \bibnamefont{and}
  \bibinfo{author}{\bibfnamefont{S.~Y.} \bibnamefont{Wang}},
  \bibinfo{journal}{Phys. Rev. C} \textbf{\bibinfo{volume}{88}},
  \bibinfo{pages}{027302} (\bibinfo{year}{2013}).

\bibitem[{\citenamefont{Bohr and Mottelson}(1975)}]{Bohr1975}
\bibinfo{author}{\bibfnamefont{A.}~\bibnamefont{Bohr}} \bibnamefont{and}
  \bibinfo{author}{\bibfnamefont{B.~R.} \bibnamefont{Mottelson}},
  \emph{\bibinfo{title}{Nuclear structure}}, vol.~\bibinfo{volume}{II}
  (\bibinfo{publisher}{Benjamin, New York}, \bibinfo{year}{1975}).

\bibitem[{\citenamefont{Chen et~al.}(2017)\citenamefont{Chen, Chen, Luo, Meng,
  and Zhang}}]{F.Q.Chen2017PRC}
\bibinfo{author}{\bibfnamefont{F.~Q.} \bibnamefont{Chen}},
  \bibinfo{author}{\bibfnamefont{Q.~B.} \bibnamefont{Chen}},
  \bibinfo{author}{\bibfnamefont{Y.~A.} \bibnamefont{Luo}},
  \bibinfo{author}{\bibfnamefont{J.}~\bibnamefont{Meng}}, \bibnamefont{and}
  \bibinfo{author}{\bibfnamefont{S.~Q.} \bibnamefont{Zhang}},
  \bibinfo{journal}{Phys. Rev. C} \textbf{\bibinfo{volume}{96}},
  \bibinfo{pages}{051303} (\bibinfo{year}{2017}).

\bibitem[{\citenamefont{Streck et~al.}(2018)\citenamefont{Streck, Chen, Kaiser,
  and Mei\ss{}ner}}]{Streck2018PRC}
\bibinfo{author}{\bibfnamefont{E.}~\bibnamefont{Streck}},
  \bibinfo{author}{\bibfnamefont{Q.~B.} \bibnamefont{Chen}},
  \bibinfo{author}{\bibfnamefont{N.}~\bibnamefont{Kaiser}}, \bibnamefont{and}
  \bibinfo{author}{\bibfnamefont{U.-G.} \bibnamefont{Mei\ss{}ner}},
  \bibinfo{journal}{Phys. Rev. C} \textbf{\bibinfo{volume}{98}},
  \bibinfo{pages}{044314} (\bibinfo{year}{2018}).

\bibitem[{\citenamefont{Ring and Schuck}(1980)}]{Ring1980book}
\bibinfo{author}{\bibfnamefont{P.}~\bibnamefont{Ring}} \bibnamefont{and}
  \bibinfo{author}{\bibfnamefont{P.}~\bibnamefont{Schuck}},
  \emph{\bibinfo{title}{The nuclear many body problem}}
  (\bibinfo{publisher}{Springer Verlag, Berlin}, \bibinfo{year}{1980}).

\bibitem[{\citenamefont{Chen et~al.}(2013)\citenamefont{Chen, Zhang, Zhao,
  Jolos, and Meng}}]{Q.B.Chen2013PRC}
\bibinfo{author}{\bibfnamefont{Q.~B.} \bibnamefont{Chen}},
  \bibinfo{author}{\bibfnamefont{S.~Q.} \bibnamefont{Zhang}},
  \bibinfo{author}{\bibfnamefont{P.~W.} \bibnamefont{Zhao}},
  \bibinfo{author}{\bibfnamefont{R.~V.} \bibnamefont{Jolos}}, \bibnamefont{and}
  \bibinfo{author}{\bibfnamefont{J.}~\bibnamefont{Meng}},
  \bibinfo{journal}{Phys. Rev. C} \textbf{\bibinfo{volume}{87}},
  \bibinfo{pages}{024314} (\bibinfo{year}{2013}).

\bibitem[{\citenamefont{Chen et~al.}(2016)\citenamefont{Chen, Zhang, Zhao,
  Jolos, and Meng}}]{Q.B.Chen2016PRC}
\bibinfo{author}{\bibfnamefont{Q.~B.} \bibnamefont{Chen}},
  \bibinfo{author}{\bibfnamefont{S.~Q.} \bibnamefont{Zhang}},
  \bibinfo{author}{\bibfnamefont{P.~W.} \bibnamefont{Zhao}},
  \bibinfo{author}{\bibfnamefont{R.~V.} \bibnamefont{Jolos}}, \bibnamefont{and}
  \bibinfo{author}{\bibfnamefont{J.}~\bibnamefont{Meng}},
  \bibinfo{journal}{Phys. Rev. C} \textbf{\bibinfo{volume}{94}},
  \bibinfo{pages}{044301} (\bibinfo{year}{2016}).

\end{thebibliography}
\end{document}